\documentclass[11pt]{article}

\usepackage[margin=1in]{geometry}
\usepackage{amsmath,amssymb}
\usepackage{graphicx}
\usepackage{mathrsfs}
\usepackage{natbib}
\usepackage{currvita}

\bibpunct{[}{]}{;}{a}{,}{;}
\renewcommand{\Im}{\ensuremath{\operatorname{Im}}}
\renewcommand{\S}{Sec.\ }
\renewcommand{\d}[1]{\ensuremath{\operatorname{d}\!{#1}}}
\def\figref{Figure~\ref}

\def\x{\mathbf x}
\def\u{\mathbf u}
\def\U{\mathbf U}
\def\C{\mathcal C}
\def\F{\mathcal F}
\def\G{\mathcal G}
\def\H{\mathcal H}
\def\K{\mathcal K}
\def\M{\mathcal M}

\def\ILz{IL$^0$}
\def\ILo{IL$^1$}
\def\ILi{IL$^\infty$}

\begin{document}

\title{Multilayer shallow-water model with stratification and shear}

\author{F.\ J.\  Beron-Vera\\ Department of Atmospheric Sciences\\
Rosenstiel School of Marine \& Atmospheric Science\\ University of
Miami\\ Miami, FL 33145 USA\\ fberon@miami.edu} 

\date{Started 11 December 2003; this version \today.} 

\maketitle

\begin{abstract}
The purpose of this paper is to present a shallow-water-type model
with multiple inhomogeneous layers featuring variable linear velocity
vertical shear and startificaion in horizontal space and time. This
is achieved by writing the layer velocity and buoyancy fields as
linear functions of depth, with coefficients that depend arbitrarily
on horizontal position and time. The model is a generalization of
Ripa's (1995) single-layer model to an arbitrary number of layers.
Unlike models with homogeneous layers the present model is able to
represent thermodynamics processes driven by heat and freshwater
fluxes through the surface or mixing processes resulting from fluid
exchanges across contiguous layers. By contrast with inhomogeneous-layer
models with depth-independent velocity and buoyancy, the model
derived here can sustain explicitly at low frequency a current in
thermal wind balance (between the vertical vertical shear and the
horizontal density gradient) within each layer. In the absence of
external forcing and dissipation, energy, volume, mass, and buoyancy
variance constrain the dynamics; conservation of total zonal momentum
requires in addition the usual zonal symmetry of the topography and
horizontal domain.  The inviscid, unforced model admits a formulation
suggestive of a generalized Hamiltonian structure, which enables
the classical connection between symmetries and conservation laws
via Noether's theorem.  A steady solution to a system involving one
Ripa-like layer and otherwise homogeneous layers can be proved
formally (or Arnold) stable using the above invariants. A model
configuration with only one layer has been previously shown to
provide: a very good representation of the exact vertical normal
modes up to the first internal mode; an exact representation of
long-perturbation (free boundary) baroclinic instability; and a
very reasonable representation of short-perturbation (classical
Eady) baroclinic instability. Here it is shown that substantially
more accurate overall results with respect to single-layer calculations
can be achieved by considering a stack of only a few layers. A
similar behavior is found in ageostrophic (classical Stone) baroclinic
instability by describing accurately the dependence of the solutions
on the Richardson number with only two layers.


\end{abstract}

\section{Introduction}

\subsection{Motivation}

There is renewed interest to construct models for the study of the
dynamics in the upper ocean (i.e., above the main thermocline,
including the mixed layer) such that:
\begin{enumerate}
\item[1)] are capable of incorporating thermodynamic processes while
  maintaining the \emph{two-dimensional structure} of the rotating
  shallow-water equations, a paradigm of ocean dynamics on scales
  longer than a few hours \citep{Pedlosky-87}; and
\item[2)] preserve the \emph{geometric} (\emph{generalized
  Hamiltonian}) \emph{structure} of the exact three-dimensional
  models from which they derive \citep{Holm-etal-02}.
\end{enumerate}
Property 1) promises fundamental understanding of ocean processes
which are difficult---if not impossible---to be attained using ocean
general circulation models.  Property 2) enables applying a recent
flow-topology-preserving framework \citep{Holm-15} to build
parametrizations \citep{Cotter-etal-20} of unresolvable submesoscale
motions and this way investigating the contribution of these to
transport at resolvable scales, a topic of active research
\citep{McWilliams-16}.

\subsection{Background}

Back in the late 1960s and early 1970s and independently by various
authors \citep{Obrien-Reid-67, Dronkers-69, Lavoie-72}, the rotating
shallow-water model was extended by allowing for horizontal and
temporal variations of the density field, while keeping it as well
as the velocity field independent of depth.  In the simplest setting,
e.g., with one active layer floating atop an abyssal layer of inert
fluid, the resulting \emph{inhomogeneous}-layer model enables the
investigation of thermodynamic processes in the upper ocean driven
by heat and freshwater fluxes across the surface.  Due to the
two-dimensional nature of the model, the computational coast involved
in such an investigation is consideraably much lower than that
produced by an ocean general circulation model
\citep{Anderson-McCreary-85a, McCreary-etal-97}.

Following nomenclature introduced in \citet[hereinafter R95]{Ripa-JFM-95},
we will refer to the model above as \ILz,  indicating that it
represents an inhomogenous-layer model wherein fields are not allowed
to vary in the vertical.  The homogeneous-layer shallow-water model
will be called HL. Additional, more recent terminology for the \ILz{
}is ``thermal rotating shallow-water model'' \citep{Warnerford-Dellar-13,
Zeitlin-18}, which emphasizes the ability of the \ILz{ }to include
(horizontal) gradients of temperature.  The \ILz{ }is also being
called ``Ripa model'' in the literature \citep{Dellar-03,
Desveaux-etal-15, Mungkasi-Roberts-16, Sanchez-etal-16, Rehman-etal-18,
Britton-Xing-20}, in recognition of Pedro Ripa's contribution to
its understanding \citep{Ripa-GAFD-93, Ripa-Amsterdam-94, Ripa-RMF-96,
Ripa-JFM-95, Ripa-DAO-99}.  We will reserve that to refer to the
model generalized here, which was introduced in R95.

The assessment on the computational cost efficiency of the \ILz{
}holds even when more than one active layer is considered
\citep{Schopf-Cane-83, McCreary-Kundu-88, McCreary-etal-91,
McCreary-etal-01, Zavala-etal-02} or when the abyssal layer is
activated and rests over irregular topography \citep{Beier-97,
Beier-Ripa-99, Palacios-etal-00}.  Furthermore, due the simplicity
of the \ILz{ }compared to the primitive equations for arbitrarily
stratified fluid, referred to herein as \ILi, it has facilitated
conceptual understanding of basic aspects of the upper-ocean dynamics
and thermodynamics \citep{Ripa-JPO-97a, Ripa-AMS-01, Beron-Ripa-02,
Ripa-Kluwer-03}.  Due in part to this very important reason, namely,
the possibility to gain insight that is difficult to attain
with an ocean general circulation model, the \ILz{ }has been recently
revisited \citep{Gouzien-etal-17, Zeitlin-18, Lahaye-etal-20,
Holm-etal-20}.

A multilayer version of the \ILz{ }was derived in \citet{Ripa-GAFD-93}
and a low-frequency approximation was developed in \citet{Ripa-RMF-96};
cf.\ recent rederivations in \citet{Warnerford-Dellar-13, Holm-etal-20}.
The no-vertical-variation ansatz cannot be maintained under the
exact dynamics produced by the \ILi{ }when horizontal density
gradients are present.  The recipe used to keep the dynamical fields
depth independent is to vertically average the horizontal pressure
gradient.  (Some authors \citep[e.g.,][]{Fukamachi-etal-95} postulate
a turbulent momentum flux that exactly cancels the vertical variation
of the horizontal pressure gradient, but this is no more than an
ad-hoc hypothesis which sheds no light on the problem.) While this
is clearly an approximation, \citet{Ripa-GAFD-93} showed that it
does not spoil the integrals of motion and generalized Hamiltonian
structure of the problem.

Furthermore, the \ILz{ }possess a Lie--Poisson Hamiltonian structure
\citep{Dellar-03} and associated with it an Euler--Poincare variational
formulation \citep{Brocker-etal-18} wherein the Hamilton principle's
Lagrangian follows by vertically averaging that of the \ILi{
}\citep{Holm-Luesink-20}.  When the equations of motion are derived
in this formulation, there is a natural way to express three
fundamental relations \citep{Holm-etal-02}.  These are: 1) the
Kelvin circulation theorem, 2) the advection equation for potential
vorticity, and 3) an infinite family of conserved Casimir invariants
(arising from Noether's theorem for the symmetry of Eulerian fluid
quantities under Lagrangian particle relabelling).  The Euler--Poincare
formulation provides a means to consistently introduce data-driven
parameterizations of stochastic transport using the SALT (stochastic
advection by Lie transport) algorithm \citep{Holm-15, Holm-Luesink-20},
enabling data assimilation in a geometry-preserving context.

The \ILz{ }provides an attractive framework for applying the SALT
algorithm to derive parameterizations for unresolved submesoscale
motions in the upper ocean.  Indeed, numerical simulations of the
\ILz{ }\citep{Ochoa-etal-98, Pinet-Pavia-00, Gouzien-etal-17} tend
to reveal small scale circulations that resemble quite well
\citep{Holm-etal-20} submesoscale filament rollups often observed
in satellite-derived ocean color images.  Such submesoscale motions
may be unresolvable in many computational simulations.  The extent
to which they contribute to fluid transport at resolvable scales
is a subject of active investigation \citep{McWilliams-16} that the
SALT stochastic version of the \ILz{ }may cast light on.

\subsection{Limitations of the \ILz}

Despite the above geometric properties of the \ILz, it has a number
of less attractive aspects, which can be consequential for the
production of small scale circulations in the model.  Discussed in
detail by \citet{Ripa-DAO-99}, these include:
\begin{enumerate}
\item[1)] In addition to the classical Poincare and Rossby waves,
  the \ILz{ }represents variations of the thickness and density
  that do not change the vertical average of the pressure gradient
  \citep{Ripa-JFM-95, Ripa-JGR-96}.  This mode is not present in
  the \ILi.
\item[2)] A uniform flow may be unstable \citep{Fukamachi-etal-95,
  Young-Chen-95, Ripa-JGR-96}.  A priori, this phenomenon seems to
  be something different than baroclinic instability.  For instance,
  unlike Eady's problem, it experiences an ``ultraviolet
  divergence'' in the sense that a short-wave cutoff is lacking.
\item[3)] Since the dynamical fields are kept depth independent
  within each layer, there is no explicit representation of the
  thermal wind balance, between the velocity vertical shear and the
  horizontal density gradient, which dominates at low frequency.
\end{enumerate}
An important additional liminitation imposed by the depth independence
of the dynamical fields, and particularly the buoyancy, is:
\begin{enumerate}
  \item[4)] The \ILz{ } cannot represent the restratification of
  the oceanic surface mixed layer resulting from ageostrophic
  baroclinic instability of lateral density gradients, which tend
  to slump from the horizontal to the vertical \citep{Tandon-Garrett-94,
  Haine-Marshall-98, Boccaletti-etal-07}.
\end{enumerate}

\subsection{The \ILo}

To cure the unwanted features of the \ILz, R95 proposed the following
improved closure to incorporate thermodynamic processes in a one-layer
ocean model not restricted to low frequencies: 
\begin{quote} 
  \emph{in addition to allowing arbitrary velocity and buoyancy
  variations in horizontal position and time, the velocity and
  buoyancy fields are also allowed to vary linearly with depth.}
\end{quote}
Ripa's single-layer model, denoted
\ILo, enjoys a number of properties which make it very promising.
For instance:
\begin{enumerate}
\item[1)] The \ILo{ }represents explicitly the thermal wind balance
  which dominates at low frequency.
\item[2)] The free waves supported by the \ILo{ }(Poincar\'{e},
  Rossby, midlatitude coastal Kelvin, equatorial, etc.) are a very
  good approximation to the first and second vertical modes in the
  exact model with unlimited vertical variation.
\item[3)] The \ILo{ }provides an exact representation of long-perturbation
  baroclinic instability and a very reasonable representation of
  short-perturbation baroclinic instability.
\end{enumerate}

\subsection{This paper}

In this paper I present a generalization of the \ILo{ }to an arbitrary
number of layers, including two possible (mathematically equivalent)
vertical configurations (\S \ref{model}). The model obtained
incorportaes additonal flexibility to treat more complicated problems
than those that can be tackled with only one layer. With a single
layer in a reduced-gravity setting mixed-layer processes can be
minimally modeled.  Including additional layers can lead to a more
accurate representation of such processes.  On the other hand,
considering a stack of several layers atop an irregular bottom will
enable investigating the influence of the ocean's interior and even
topographic effects.  Several aspects of the gereneralized \ILo{
}are discussed in \S \ref{properties}.  These include: remarks on
submodels derived from the generalized model as special cases (\S
\ref{submodels}); the nature of the layer boundaries (\S
\ref{boundaries}); the model conservation laws (\S \ref{invariants});
a discussion on circulation theorems (\S \ref{Kelvin}); a formulation
of the model suggstive of a generalized Hamiltonian structure (\S
\ref{Hamilton}); a formal stability theorem (\S \ref{Arnold});
results on vertical normal modes (\S \ref{waves}) and on baroclinic
instability (\S \ref{instability}), both quasigeostrophic and
ageostrophic, which demonstrate that improved performance with
respect to the single-layer results can be attained by considering
only a few more layers; and the incorporation of forcing in the
model equations (\S \ref{forcing}).  Section \ref{discussion} closes
the paper with some concluding remarks.

\section{The multilayer \ILo\label{model}}

\begin{figure}[t!]
  \centering
  \includegraphics[width=.75\textwidth]{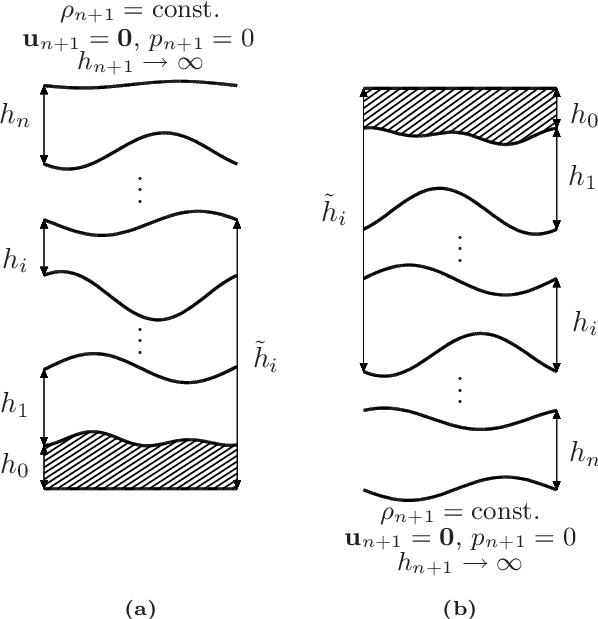}%
  \caption{The two possible vertical configurations of the $n$-\ILo\,
  are rigid bottom \textbf{(a)} and rigid lid \textbf{(b)}.  Within
  each layer the velocity and buoyancy fields not only vary arbitrarily
  with the horizontal position and time, but also linearly with depth.}
  \label{fig:VertConf}%
\end{figure}

Consider a stack of $n$ active fluid layers with thickness $h_i(\x,t),$
$i = 1,\dotsc , n$, where $\x$ is horizontal position and $t$ sands
for time (\figref{fig:VertConf}). The geometry can be either
planar or spherical; in the former case the vertical coordinate,
$z$, is perpendicular to the plane, whereas in the latter it is
radial. The total thickness is $h(\x,t) = \sum_j h_j(\x,t).$ The
stack of inhomogeneous-density layers can be either limited from
below by a rigid bottom, $z = h_0(\x)$, or from above by a rigid
lid, $z = -h_0(\x).$ The usual choice in the rigid lid case is $h_0
\equiv 0$; however, laboratory experiments are often designed to
have a nonhorizontal top lid. The remaining boundary in the
rigid-bottom (resp., rigid-lid) configuration is a soft interface
with a passive, infinitely thick layer of lighter (resp., denser)
homogeneous fluid of density $\rho_{n+1}$. Although vacuum
($\rho_{n+1}\equiv 0$) is the typical setting in the rigid-bottom
configuration, the choice $\rho_{n+1}\neq 0$ can be useful to study
of deep flows over topography.

\begin{figure}
 \centering%
 \includegraphics[width=.75\textwidth]{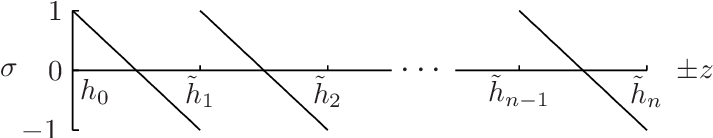}%
 \caption{Vertical coordinate choice. Within each layer the rescaled
 vertical coordinate $\sigma$ varies linearly from $\pm 1$, at the
 base, to $\mp1$, at the top. The upper (resp., lower) sign corresponds
 to the rigid-bottom (resp., rigid-lid) configuration of
 \figref{fig:VertConf}.}
 \label{fig:VertCoord}%
\end{figure}

A key element to generalize Ripa's model is to define a scaled vertical
coordinate $\sigma_i$ so as to vary linearly from $\pm1$ at the base to 
$\mp 1$ at the top of the $i$th layer (\figref{fig:VertCoord}): 
\begin{equation}
  \pm z=:\tilde{h}_{i-1}(\x,t)+\frac{1-\sigma_i}{2}h_{i}(\x,t) =
  \nu_{i}(\x,\sigma_i,t), \label{sigma}
\end{equation}
where 
\begin{equation}
  \tilde h_i(\x,t) := h_0(\x) + \sum_{j=1}^i h_j(\x,t)
\end{equation}
[henceforth an upper (resp., lower) sign will correspond to the
rigid-bottom (resp., rigid-lid) configuration]. The scaled vertical
coordinate $\sigma$ defined in R95 according to
\begin{equation}
  z =: h_0(\x) + \frac{1}{2}(\sigma -1)h(\x,t) = \nu (\x,\sigma,t)
\end{equation}
relates to the $i$th-layer scaled vertical coordinate $\sigma_i$
defined here through
\begin{equation}
  \sigma = 1 - 2\sum_{j=1}^{i-1} \frac{h_j}{h} +
  (1 - \sigma_i)\frac{h_i}{h}.
\end{equation}

Let an overbar denote vertical average within the $i$th layer:
\begin{equation}
  \bar a_i(\x,t) := \frac{1}{2}\int_{-1}^{+1} a(\x,\sigma,t)
  \d{\sigma_i} = \frac{1}{2}\int_{-1}^{+1} a_i(\x,\sigma_i,t)
  \d{\sigma_i}.
\end{equation}
Following R95 closely, the $i$th-layer horizontal velocity and
buoyancy fields are written, respectively, as
\begin{subequations}
\label{ansatz}
\begin{align}
  \u_{i}(\x,\sigma _{i},t) &= \mathbf{\bar{u}}_{i}(\x ,t)+\sigma
  _{i}\u_{i}^{\sigma }(\x,t),\\ \vartheta _{i}(\x,\sigma _{i},t)
  &= \bar{\vartheta}_{i}(\x ,t)+\sigma _{i}\vartheta _{i}^{\sigma}(\x,t),
\end{align}
which can be regarded as a truncation of an expansion in orthogonal
polynomials of $\sigma _{i}$ of the form 
\end{subequations}
\begin{equation}
  a_{i}(\x,\sigma _{i},t) = \bar{a}_{i}(\x,t) +
  \sigma_{i}a_{i}^{\sigma}(\x,t) +
  \tfrac{1}{2}\left(\sigma_{i}^{2}-\tfrac{1}{3}\right) a_{i}^{\sigma
  \sigma}(\x,t) + \tfrac{1}{6}\left(\sigma_{i}^{3} -
  \tfrac{3}{5}\sigma_{i}\right) a_{i}^{\sigma\sigma\sigma}(\x,t) +
  \dotsc,
\end{equation}
where $a_{i}^{\sigma}:=\overline{\partial_{\sigma_{i}}a_{i}},$ $%
a_{i}^{\sigma\sigma}:=\overline{\partial_{\sigma_{i}\sigma_{i}}a_{i}},$
etc. \citep[][]{Ripa-DAO-99}. The $i$th-layer buoyancy is defined
as
\begin{equation}
  \vartheta_{i}(\x,\sigma _{i},t) := \pm g\,\frac{\rho_{i}(\x,\sigma_{i},t)
  - \rho_{n+1}}{\rho_\mathrm{r}},
\end{equation}
where the upper (resp., lower) sign corresponds to the rigid-bottom
(resp., rigid-lid). Here, $g$ is gravity, $\rho _{i}(\x,\sigma_{i},t)
= \bar{\rho}_{i}(\x,t) + \sigma _{i}\rho _{i}^{\sigma }(\x,t)$ is
the (variable) density in the $i$th layer, and $\rho_{\mathrm{r}}$
denotes the (constant) reference density used in the Boussinesq
approximation.  Physically admissible buoyancy values, i.e., everywhere
positive and monotonically increasing (resp., decreasing) with depth
in the rigid-bottom (resp., rig-lid) case, are such that
\begin{equation}
\bar{\vartheta}_{i}>\vartheta _{i}^{\sigma }>0,\quad \bar{\vartheta}_{i}-%
\bar{\vartheta}_{i+1}\geq \vartheta _{i}^{\sigma }+\vartheta _{i+1}^{\sigma
}.  \label{CondTheta}
\end{equation}%
If $n_{i}^{2}(\x,t) > 0$ is the square of the instantaneous
Brunt-V\"{a}is\"{a}l\"{a} frequency within the $i$th layer, then
note that
\begin{equation}
  \vartheta_{i}^{\sigma} = \tfrac{1}{2}n_{i}^{2}h_{i}.
\end{equation}
In order to obtain the equations for the $n$-layer version of Ripa's model
one must proceed as follows:

\begin{enumerate}
\item[1)] Substitute ansatz (\ref{ansatz}) in the inviscid, unforced,
\emph{primitive equations} (namely, rotating, incompressible,
hydrostatic, Euler--Boussinesq equations) for arbitrarily stratified
fluid (\ILi), which can be written as
\begin{subequations}
\label{ILinf}
\begin{gather}
  \mathrm{D}\vartheta = 0, \\
  \left. \partial _{t}\right| _{\sigma }h+\left. \mathbf{\nabla }\right|
  _{\sigma }\cdot h\u+h\partial _{\sigma }\mu =0, \\
  \mathrm{D}\u+f\mathbf{\hat{z}}\times \u+\left. \mathbf{
  \nabla }\right| _{\sigma }p+\vartheta \left. \mathbf{\nabla }\right|
  _{\sigma }\nu =\mathbf{0}, \\
  \partial _{\sigma }p-\tfrac{1}{2}h\vartheta = 0,
\end{gather}
where
\begin{equation}
  \mu :=\mathrm{D}\sigma =\frac{2\mathrm{D}h_{0}+\left( 1-\sigma
  \right) \mathrm{D}h\mp 2w}{h}.
\end{equation}
In (\ref{ILinf}), 
\begin{equation}
  \mathrm{D} := \partial_t\vert_\sigma + \u \cdot \nabla\vert_\sigma
  + \mu \partial_\sigma
\end{equation}
is the material derivative, where $\left. \partial _{t}\right|
_{\sigma }$ and $\left. \mathbf{\nabla }\right| _{\sigma }$ indicate,
respectively, that the partial time derivative and the horizontal
gradient operate at constant $ \sigma $ [note that $\left. \partial
_{t}\right| _{\sigma }a\equiv \partial _{t}a$ and $\left. \mathbf{\nabla
}\right| _{\sigma }a\equiv \mathbf{\nabla } a$, and thus
$\mathrm{D}a\equiv \partial _{t}a+\u\cdot \mathbf{ \nabla }a,$ for
any $a(\x,t)$]; $f$ is the Coriolis parameter (twice the local
angular rotation frequency); and $\mathbf{\hat{z}}$ is the vertical
unit vector. Also in (\ref{ILinf}), $(\u,w)$ is the three-dimensional
velocity, $\mu$ denotes the $\sigma$-vertical velocity, $\vartheta$
stands for buoyancy, and $p$ is a \emph{kinematic} pressure; the vertical
variation in all these fields is unrestricted. Equations (\ref
{ILinf}a--d) are defined in $-1<\sigma <+1$ (i.e., $h_{0} < \pm z
< h_{0}+h$) and are subject to the boundary conditions
\begin{align}
  \mu &= 0\text{\quad at}\quad \sigma = \{-1,+1\},\\
  p &= 0\text{\quad at\quad }\sigma = -1.
\end{align}
\end{subequations}
Note that boundary conditions (\ref{ILinf}g) can be expressed as
$(\left. \partial _{t}\right| _{\sigma }+\u\cdot \left. \mathbf{
\nabla}\right| _{\sigma })(h_{0}+\frac{1}{2}[1\mp 1]h\mp \zeta )=0$
at the base of the layer and $(\left. \partial _{t}\right| _{\sigma
}+\u \cdot \left. \mathbf{\nabla }\right| _{\sigma
})(h_{0}+\frac{1}{2}[1\pm 1])h\mp \zeta )=0$ at the top of the
layer. Here, $\zeta (\x,\sigma ,t)$ is the vertical displacement
of a constant-density surface or isopycnal, which, by virtue
(\ref{ILinf}a), relates to the vertical velocity through
$w=\mathrm{D}\zeta $. These conditions thus indicate that a fluid
particle initially on a given boundary remains there at all times
conserving its density. A particular case is one in which all
particles on the boundary have the same density, i.e.,
$h_{0}+\frac{1}{2}[1\mp 1]h\mp \zeta =\mathrm{const}$ at the base
of the layer and/or $h_{0}+\frac{1}{2} [1\pm 1]h\mp \zeta
=\mathrm{const}$ at the top of the layer.

\item[2)] Replace all occurrences of $\sigma ^{2}$ by its vertical
average (i.e., $\sigma ^{2}\mapsto \frac{1}{3}$) to preserve the linear
vertical structure within each layer.

\item[3)] Collect terms in powers of $\sigma $ and equate them to zero
afterwards.
\end{enumerate}

\noindent The equations that result from the above three-step
procedure constitute the $n$-\ILo, and are given by:
\begin{subequations}\label{IL1n}
\begin{gather}
  \overline{\mathrm{D}_{i}\vartheta _{i}}=0, \\
  (\mathrm{D}_{i}\vartheta _{i})^{\sigma }=0, \\
  \partial _{t}h_{i}+\mathbf{\nabla }\cdot h_{i}\mathbf{\bar{u}}_{i}=0, \\
  \overline{\mathrm{D}_{i}\u_{i}}+f\mathbf{\hat{z}}\times \mathbf{\bar{
  u}}_{i}+\overline{\mathbf{\nabla }p}_{i}=\mathbf{0}, \\
  (\mathrm{D}_{i}\u_{i})^{\sigma }+f\mathbf{\hat{z}}\times \u
  _{i}^{\sigma }+(\mathbf{\nabla }p_{i})^{\sigma }=\mathbf{0}.
\end{gather}
Here, 
\begin{align}
  \overline{\mathrm{D}_{i}a_{i}} &= \partial _{t}\bar{a}_{i}+\mathbf{\bar{u}}
  _{i}\cdot \mathbf{\nabla }\bar{a}_{i}+\tfrac{1}{3}h_{i}^{-1}\mathbf{\nabla }
  \cdot h_{i}a_{i}^{\sigma }\u_{i}^{\sigma }, \\
  (\mathrm{D}_{i}a_{i})^{\sigma } &= \partial _{t}a_{i}^{\sigma }+\mathbf{\bar{
  u}}_{i}\cdot \mathbf{\nabla }a_{i}^{\sigma }+\u_{i}^{\sigma }\cdot 
  \mathbf{\nabla }\bar{a}_{i},
\end{align}
are the mean and $\sigma $ components of the material derivative of any
field $a_{i}(\x,\sigma _{i},t)=\bar{a}_{i}(\x,t)+\sigma
_{i}a_{i}^{\sigma }(\x,t)$ in the $i$th layer; and
\begin{align}
  \overline{\mathbf{\nabla }p}_{i} &= (\bar{\vartheta}_{i}-\tfrac{1}{3}
  \vartheta _{i}^{\sigma })\mathbf{\nabla }h_{i} +
  \tfrac{1}{2}h_{i}\mathbf{\nabla }(\bar{\vartheta}_{i}-\tfrac{1}{3}
  \vartheta _{i}^{\sigma }) + \bar{\vartheta}_{i} \mathbf{\nabla
  }\tilde{h}_{i-1} +\mathbf{ \nabla }\sum\limits_{j=i+1}^{n}h_{j}\bar{
  \vartheta}_{j},\\ 
  \left( \mathbf{\nabla }p_{i}\right) ^{\sigma }
  &= \tfrac{1}{2}\vartheta _{i}^{\sigma }\mathbf{\nabla
  }h_{i}+\tfrac{1}{2}h_{i}\mathbf{\nabla }\bar{ \vartheta}_{i}+\vartheta
  _{i}^{\sigma }\mathbf{\nabla }\tilde{h}_{i-1},
\end{align}
\end{subequations}
which are the mean and $\sigma $ components of the $i$th-layer pressure
gradient force.

System (\ref{IL1n}) consists of $7n$ evolution equations in the
$7n$ independent fields $(\bar{\vartheta}_{i},\vartheta _{i}^{\sigma
},h_{i},\allowbreak \mathbf{\bar{u}}_{i},\u_{i}^{\sigma })$, $
i=1,\cdots ,n.$ The coupling among different layer quantities is
provided by the last terms on the right hand side of the pressure
forces (\ref{IL1n}h,i). It is important to note that
the dynamics in both the rigid-bottom and rigid-lid configurations
is described by system (\ref{IL1n}); no double signs are needed.
The latter must be taken into account, however, in the computation
of the \emph{total} pressure in the $i$th layer, which, up to the addition
of an irrelevant constant, is given by $\rho_{\mathrm{r} }p_{i}\pm
\rho _{n+1}gz$, where
\begin{equation}
  p_{i}=\tfrac{1}{2}\left( 1+\sigma _{i}\right) h_{i}\bar{\vartheta}_{i}-
  \tfrac{1}{4}(1-\sigma _{i}^{2})h_{i}\vartheta _{i}^{\sigma
  }+\sum\limits_{j=i+1}^{n}h_{j}\bar{\vartheta}_{j}.
\end{equation}

Finally, equations (\ref{IL1n}) are satisfied in some closed but
multiply-connected horizontal domain, say $D.$ On $\partial D,$ i.e., the
union of each disconnected part of the solid boundary of $D$, the zero
normal flow condition holds:
\begin{equation}
  \mathbf{\bar{u}}_{i}\cdot \mathbf{\hat{n}}=0=\u_{i}^{\sigma }\cdot 
  \mathbf{\hat{n}}\quad \text{on}\quad \partial D  \label{BC}
\end{equation}
where $\mathbf{\hat{n}}$ is normal to $\partial D.$

\section{Discussion of several aspects of the $n$-\ILo\label{properties}}

\subsection{Submodels\label{submodels}}

Any initial state with uniform buoyancy inside each layer ($\bar{\vartheta}%
_{i}=\mathrm{const}$ and $\vartheta _{i}^{\sigma }\equiv 0$) and
vanishing vertical shear ($\u_{i}^{\sigma }\equiv 0$) is readily
seen to be preserved by (\ref{IL1n}); consequently, the $n$-HL (a
model with $n$ homogeneous layers) follows from (\ref{IL1n}) as a
particular case, just as it does it from the (exact, three-dimensional)
\ILi{ }model (\ref{ILinf}). In other words, the $n$-HL evolves on
an invariant submanifold of both the $n$-\ILo{ }and \ILi.  Noteworthy,
the $n$-HL is exact for a stepwise density stratification; however,
as mentioned above, it is not able to accommodate thermodynamic
processes, e.g., due to heat and buoyancy fluxes across the ocean
surface. The $n$-\ILz{ }developed in \citet[]{Ripa-GAFD-93} follows
from (\ref{IL1n}) upon neglecting $\u_{i}^{\sigma }$ and $\vartheta
_{i}^{\sigma }$; note that an initial condition with $\u_{i}^{\sigma
}\equiv 0$ and $\vartheta_{i}^{\sigma }\equiv 0$ is preserved neither
by (\ref{IL1n}) nor by (\ref{ILinf}), so the $n$-\ILz{ }is not a
particular solution of neither the $n$-\ILo{ }nor the \ILi. Ignoring
$\u_{i}^{\sigma }$ in (\ref{IL1n}) results in a model with
$\u_{i}^{\sigma }\equiv 0$ but $\vartheta _{i}^{\sigma }\neq 0$
which provides a generalization for \citeauthor{Schopf-Cane-83}'s
[1983] intermediate layer model.  Alternatively, omission of
$\vartheta _{i}^{\sigma }$ in system (\ref{IL1n}) gives a model
with $\vartheta _{i}^{\sigma }\equiv 0$ but $\u_{i}^{\sigma }\neq
0$. This model differs from earlier related models \citep[][]{Benilov-93,
Young-94, Scott-Willmott-02} in that it is not restricted to
low-frequency motions and that it explicitly represents vertical
shear within each of an arbitrary number of layers.

\subsection{Layer boundaries\label{boundaries}}

Consistent with ansatz (\ref{ansatz}) and the assumption of zero
mass transport across layer boundaries, the $\sigma $-vertical
velocity (\ref {ILinf}e) in the $i$th layer reads
\begin{equation}
  \mu _{i}=\frac{1-\sigma _{i}^{2}}{2h_{i}}\,\mathbf{\nabla }\cdot
  h_{i}\u_{i}^{\sigma },  
  \label{mu}
\end{equation}
which vanishes at the base and the top of the layer. Consequently,
$ (\partial _{t}+[\mathbf{\bar{u}}_{i}\mp \u_{i}^{\sigma }]\cdot
\mathbf{\nabla })(\tilde{h}_{i-1}+\frac{1}{2}[1\mp 1]h_{i}\mp \lbrack
\bar{ \zeta}_{i}\mp \zeta _{i}^{\sigma }])=0$ at the base of the
$i$th layer and $ (\partial _{t}+[\mathbf{\bar{u}}_{i}\pm \u_{i}^{\sigma
}]\cdot \mathbf{\nabla })(\tilde{h}_{i-1}+\frac{1}{2}[1\pm 1]h_{i}\mp
\lbrack \bar{ \zeta}_{i}\pm \zeta _{i}^{\sigma }])=0$ at the top
of the $i$th layer.  Namely, the layer boundaries (interfaces and
rigid bottom or lid) of the $n$-\ILo{ } are material surfaces on
which each fluid particle retains its density. This includes the
particular situation in which all fluid particles on these boundaries
have the same density, i.e., $\tilde{h}_{i-1}+\frac{1}{2}[1\mp
1]h_{i}\mp \lbrack \bar{\zeta}_{i}\mp \zeta _{i}^{\sigma
}]=\mathrm{const}$ at the base of the $i$th layer and $
\tilde{h}_{i-1}+\frac{1}{2}[1\pm 1]h_{i}\mp \lbrack \bar{\zeta}_{i}\pm
\zeta _{i}^{\sigma }]=\mathrm{const}$ at the top of the $i$th layer.
The latter situation, which is most likely to happen far away from
the ocean surface, cannot be described by the \ILo{ }with only
one layer.

\subsection{Conservation laws\label{invariants}}

In a closed horizontal domain, on whose boundary conditions (\ref{BC}) are
satisfied, conservation of the $i$th-layer volume, mass, and buoyancy
variance is enforced, respectively, because of (\ref{IL1n}c),
\begin{equation}
  \partial _{t}\left( h_{i}\bar{\vartheta}_{i}\right) +\mathbf{\nabla }\cdot
  h_{i}(\bar{\vartheta}_{i}\mathbf{\bar{u}}_{i}+\tfrac{1}{3}\vartheta
  _{i}^{\sigma }\u_{i}^{\sigma })=0,
\end{equation}
and
\begin{equation}
  \partial _{t}(h_{i}\overline{\vartheta _{i}^{2}})+\mathbf{\nabla }\cdot
  h_{i}(\overline{\vartheta _{i}^{2}}\mathbf{\bar{u}}_{i}+\tfrac{2}{3}\bar{
  \vartheta}_{i}\vartheta _{i}^{\sigma }\u_{i}^{\sigma })=0.
\end{equation}

The total energy (sum of the energies in each layer) is also preserved in a
closed horizontal domain since
\begin{subequations}
\begin{equation}
  \partial _{t}\sum_{j}E_{j}+\mathbf{\nabla }\cdot \sum_{j}h_{j}(\bar{b}_{j}
  \mathbf{\bar{u}}_{j}+\tfrac{1}{3}b_{j}^{\sigma }\u_{j}^{\sigma })=0,
\end{equation}
where
\begin{equation}
  E_{i}:=\tfrac{1}{2}h_{i}\mathbf{\bar{u}}_{i}^{2}+\tfrac{1}{6}h_{i}(\u
  _{i}^{\sigma })^{2}+\tfrac{1}{2}h_{i}^{2}(\bar{\vartheta}_{i}-\tfrac{1}{3}
  \vartheta _{i}^{\sigma })+h_{i}\tilde{h}_{i-1}\bar{\vartheta}_{i},
  \label{eq:Ej}
\end{equation}
and
\begin{align}
  \bar{b}_{i} &:= \tfrac{1}{2}\mathbf{\bar{u}}_{i}^{2}+\tfrac{1
  }{6}(\u_{i}^{\sigma })^{2}+h_{i}(\bar{\vartheta}_{i}-\tfrac{1}{3}
  \vartheta _{i}^{\sigma })+\tilde{h}_{i-1}\bar{\vartheta}_{i}+
  \sum_{j=i+1}^{n}h_{j}\bar{\vartheta}_{j}, \\
  b_{i}^{\sigma } &:= \mathbf{\bar{u}}_{i}\cdot \u
  _{i}^{\sigma }+(\tilde{h}_{i-1}+\tfrac{1}{2}h_{i})\vartheta _{i}^{\sigma },
\end{align}
\end{subequations}
which are the mean and $\sigma $ components of the $i$th-layer Bernoulli
head. The above result follows upon realizing that $\sum_{j=1}^{n}h_{j}\bar{
\vartheta}_{j}\partial _{t}\tilde{h}_{j-1}-\sum_{j=1}^{n}\partial
_{t}h_{j}\sum_{k=j+1}^{n}\allowbreak h_{k}\bar{\vartheta}_{k}\equiv 0$, and
is largely facilitated by rewriting (\ref{IL1n}d,e) in the form
\begin{subequations}
  \label{B}
\begin{align}
  \partial _{t}\mathbf{\bar{u}}_{i}+\bar{\mu}_{i}\u_{i}^{\sigma }+h_{i}
  \mathbf{\hat{z}}\times (\bar{q}_{i}\mathbf{\bar{u}}_{i}+\tfrac{1}{3}
  q_{i}^{\sigma }\u_{i}^{\sigma })+\mathbf{\nabla }\bar{b}_{i} &=
  \mathbf{\bar{R}}_{i}, \\
  \partial _{t}\u_{i}^{\sigma }+h_{i}\mathbf{\hat{z}}\times
  (q_{i}^{\sigma }\mathbf{\bar{u}}_{i}+\bar{q}_{i}\u_{i}^{\sigma })+
\mathbf{\nabla }b_{i}^{\sigma } &=\mathbf{R}_{i}^{\sigma }.
\end{align}
\end{subequations}
Here, 
\begin{equation}
  \bar{\mu}_{i}=\tfrac{1}{3}h_{i}^{-1}\mathbf{\nabla }\cdot h_{i}\u
  _{i}^{\sigma }
\end{equation}
is the vertical average of the $i$th-layer $\sigma $-vertical velocity (\ref
{mu}); 
\begin{equation}
  \bar{q}_{i}:=h_{i}^{-1}\left( f+\mathbf{\nabla }\cdot \mathbf{\bar{u}}
  _{i}\times \mathbf{\hat{z}}\right),\quad 
  q_{i}^{\sigma }:=h_{i}^{-1}\mathbf{
\nabla }\cdot \u_{i}^{\sigma }\times \mathbf{\hat{z}}
\end{equation}
are the mean and $\sigma $ components of the $i$th-layer $\sigma
$-potential vorticity;\footnote{For a general scalar $s,$ the
$s$\emph{-potential vorticity} is defined by $ \mathcal{L}s,$ where
$\mathcal{L:}=\sum_{a}q^{a}\partial (\cdot )/\partial x^{a}$ is a
coordinate-independent representation [R95] of the Ertel operator
\citep[cf.][]{Pedlosky-87}. Here, $x^{a}$ is any coordinate and
$q^{a}=\mathcal{L}x^{a}$ is the $a$ th-component of the absolute
vorticity. Consistent with the dynamics represented by the \ILi{
}in $(\x,\sigma )$ coordinates (\ref{ILinf}), $\mathcal{L}=\mathbf{q}\cdot
\left. \mathbf{ \nabla }\right\vert _{\sigma }+q\partial _{\sigma
}$ where $\mathbf{q} :=h^{-1}\mathbf{\hat{z}}\times \partial _{\sigma
}\u$ and $ q:=h^{-1}(f+\mathbf{\hat{z}}\cdot \left. \mathbf{\nabla
}\right\vert _{\sigma }\times \u).$} and
\begin{subequations}
\begin{align}
  \mathbf{\bar{R}}_{i} &:= \tilde{h}_{i-1}\mathbf{\nabla }\bar{
  \vartheta}_{i}+\tfrac{1}{2}h_{i}\mathbf{\nabla (}\bar{\vartheta}_{i}-\tfrac{1
  }{3}\vartheta _{i}^{\sigma }), \\
  \mathbf{R}_{i}^{\sigma } &:= (\tilde{h}_{i-1}+\tfrac{1}{2}
  h_{i})\mathbf{\nabla }\vartheta _{i}^{\sigma }-\tfrac{1}{2}h_{i}\mathbf{
  \nabla }\bar{\vartheta}_{i},
\end{align}
\end{subequations}
which are rotational forces that arise as a consequence of the buoyancy
inhomogeneities within each layer ($\mathbf{\nabla }\bar{\vartheta}_{i}\neq 
\mathbf{0}\neq \mathbf{\nabla }\vartheta _{i}^{\sigma }$).

In turn, the local conservation law for the sum of the zonal momenta within
each layer is given by
\begin{subequations}
\begin{equation}
  \partial _{t}\sum_{j}M_{j}+\mathbf{\nabla }\cdot \sum_{j}\mathbf{F}
  _{j}^{M}+\partial _{x}h_{0}\sum_{j}h_{j}\bar{\vartheta}_{j}=0,
\end{equation}
where 
\begin{equation}
  \mathbf{F}_{i}^{M}:=M_{i}\mathbf{\bar{u}}_{i}+\tfrac{1}{3}h_{i}u_{i}^{\sigma
  }\u_{i}^{\sigma }+\tfrac{1}{2}\gamma h_{i}^{2}(\bar{\vartheta}_{i}-
  \tfrac{1}{3}\vartheta _{i}^{\sigma })\mathbf{\hat{x}}+\gamma h_{i+1}\bar{
  \vartheta}_{i+1}\sum_{j=1}^{i-1}h_{j}\mathbf{\hat{x}}
\end{equation}
\end{subequations}
with $u_{i}$ denoting the zonal component of $\u_{i}$ and $\mathbf{
\hat{x}}$ the unit vector in the same direction.\footnote{ The term
$-\frac{1}{6}\partial _{x}(h\vartheta _{\sigma })$ is missing on
the right-hand side of (4.6) in R95.} The above result follows
upon multiplying by $\gamma h_{i}$ the zonal component of
(\ref{IL1n}d),
\begin{equation}
  \partial _{t}\bar{u}_{i}+\mathbf{\bar{u}}_{i}\cdot \mathbf{\nabla }\bar{u}
  _{i}+\tfrac{1}{3}h_{i}^{-1}\mathbf{\nabla }\cdot h_{i}u_{i}^{\sigma }\mathbf{
  u}_{i}^{\sigma }-\left( f+\tau \bar{u}_{i}\right) \bar{v}_{i}+\gamma ^{-1}
  \overline{\partial _{x}p}_{i}=0,
\end{equation}
and realizing that $\sum_{j}h_{j}\bar{\vartheta}_{j}\partial
_{x}(\tilde{h} _{j-1}-h_{0})+\sum_{j}h_{j}\partial
_{x}\sum_{k=j+1}^{n}h_{k}\bar{\vartheta} _{k}\equiv \partial
_{x}(\sum_{j}h_{j+1}\allowbreak \bar{\vartheta} _{j+1}\allowbreak
\sum_{k=1}^{j-1}h_{k})$. At this point it is crucial to specify
whether the geometry is flat or spherical. On the sphere, $\mathbf{
\nabla }a=\left( \gamma ^{-1}\partial _{x}a,\partial _{y}a\right)
,$ for any scalar $a(\x),$ and $\mathbf{\nabla }\cdot \mathbf{a}=\gamma
^{-1}[\partial _{x}a+\partial _{y}\left( \gamma b\right) ],$ for
any vector $ \mathbf{a}=(a,b),$ where $x=(\lambda -\lambda _{0})\cos
\theta \,R$ and $ y=(\theta -\theta _{0})R$ are, respectively,
rescaled geographic longitude and latitude on the surface of the
Earth whose mean radius is $R$; and $ \gamma (y):=\cos \theta
_{0}\cos \theta $ and $\tau (y):=R^{-1}\tan \theta \equiv -\gamma
^{-1}\mathrm{d}\gamma /\mathrm{d}y$ are coefficients that characterize
the geometry of the space (the arclength element square and area
element are $\mathrm{d}\x^{2}=\gamma ^{2}\mathrm{d}x^{2}+
\mathrm{d}y^{2}$ and $\mathrm{d}^{2}\x=\gamma \mathrm{d}x\mathrm{d}y$
, respectively). The $i$th zonal momentum (angular momentum around
the Earth's axis) is then given by
\begin{equation}
  M_{i}:=h_{i}[\gamma \bar{u}_{i}-\Omega R(\cos \vartheta _{0}-\gamma \cos
\vartheta )],
\end{equation}
where $\Omega $ is the Earth's angular rotation rate. In the classical
$ \beta $ plane, $\gamma =1$ and $\tau =0$ so that all the operators
are Cartesian and $M_{i}=h_{i}(\bar{u}_{i}-f_{0}y-\frac{1}{2}\beta
y^{2}).$ However, the geometry in a consistent $\beta $ plane cannot
be Cartesian; instead $\gamma =1-\tau _{0}y$, $\tau =\tau _{0}/\gamma
,$ and $ M_{i}=h_{i}[\gamma \bar{u}_{i}-f_{0}y-\frac{1}{2}\beta
(1-R^{2}\tau _{0}^{2})y^{2}]$ \citep[][]{Ripa-JPO-97a}. Finally,
conservation of the total zonal momentum (sum over all layers) in
a horizontal domain in addition requires, in all cases, that both
the topography and coasts to be zonally symmetric.

\subsection{Circulation theorems\label{Kelvin}}

In the \ILi{ }the circulation of $\u+\u_{f}$, where $\mathbf{\hat
z}\cdot \mathbf\nabla\times \u_f := f$, around a material loop is
constant in time if the latter is chosen to lie on an isopycnic
surface.\footnote{Under this condition the circulation of $\u$ is
not preserved as claimed in R95.} This is known as the Kelvin
circulation theorem, which via Stokes' theorem implies conservation
of $\vartheta $ -potential vorticity. From the Hamiltonian mechanics
side, the Kelvin theorem is the geometrical statement of invariance
of the fluid action integral on level surfaces of $\vartheta $
\citep[e.g.,][]{Holm-96}. Existence of a Kelvin circulation property
is thus closely related to existence of a (constrained) Hamilton's
principle for the \ILi. The $n$-\ILo{ }does not hold such a circulation
property. As a consequence, the evolution of the $i$th-layer
$\vartheta $-potential vorticity is not correctly represented. In
R95 it is shown that this is the result of the lack of information
on the vertical curvature of the horizontal velocity field. It is
easy to show, however, that the evolution of the three components
of the vorticity field are correctly represented, and, consistent
with the \ILi, neither $\bar q_i$ nor $q_i^\sigma$ are conserved.
The evolution equations of the latter fields and the horizontal
vorticity are given by equation (4.21) in R95 evaluted in the $i$th
layer (note that evaluation of $\nu$ in the $i$th layer does not
simply mean replacing $h$ by $h_{i}$).  Nonexistence of a Kelvin
circulation property for the $n$-\ILo{ }suggestes that finding a
Hamilton's principle for it is, at least, nontrivial. The $n$-\ILo{
}is nonetheless shown in \S \ref{Hamilton} to admit a formulation
suggestive of a generalized Hamiltonian structure. The $n$-\ILz,
surprisingly, possesses a Kelvin circulation property since
$\frac{\d{}}{\d{t}} \oint_{\ell
_{t}(\mathbf{\bar{u}}_{i})}(\mathbf{\bar{u}}_{i} + \u_{f}) \cdot
\d{\x} =\oint_{\ell _{t}(\mathbf{\bar{u}} _{i})}
(\tilde{h}_{i-1}\mathbf{\nabla }\bar{ \vartheta}_{i} +
\frac{1}{2}h_{i}\mathbf{\nabla }\bar{\vartheta}_{i}) \cdot \d{\x}$
holds in that model and the material loop $\ell _{t}(\mathbf{\bar{u}}_{i})$
can be chosen to lie on an isopycnic surface. Consistent with the
presence of this property, \citet[]{Dellar-03} showed that the \ILz{
}has a Lie--Poisson Hamiltonian structure which implies an analogous
Euler--Poincare variational formulation \citep[][]{Holm-etal-02}
and, hence, the existence of a Lagrangian functional.

In the $n$-\ILo{ }the following circulations theorems hold:
$\frac{\d{}}{\d{t}} \oint_{\partial D} \mathbf{ \bar{u}}_{i} \cdot
\d{\x} = \oint_{\partial D} (\mathbf{\bar{R}}_{i} -
\bar{\mu}_{i}\u_{i}^{\sigma}) \cdot \d{\x}$ and $\frac{\d{}}{\d{t}}
\oint_{\partial D} \u _{i}^{\sigma} \cdot \d{\x} = \oint_{\partial
D} \mathbf{R} _{i}^{\sigma} \cdot \d{\x}$. This contrasts with the
\ILi{ }for which the circulation of $\u$ around $\partial D$ is
time independent. Note that the circulation of $\u_{i}^{\sigma }$
around $\partial D$ would be invariant if both $\bar{\vartheta}_{i}$
and $\vartheta _{i}^{\sigma}$ were chosen such that $\mathbf{\hat{n}}\times
\mathbf{\nabla }\bar{\vartheta}_{i}=\mathbf{0=\hat{n}}\times
\mathbf{\nabla }\vartheta _{i}^{\sigma }$ on $\partial D$.\footnote{The
circulation of $\mathbf{\bar{u}}_{i}$ would not be constant in time
under these conditions as argued in R95.} However, the latter
boundary is not preserved by the $n$\ILo{ }dynamics. In opposition,
the condition $ \mathbf{\hat{n}}\times \mathbf{\nabla
}\bar{\vartheta}_{i}=\mathbf{0}$ on $ \partial D$ is preserved by
the $n$-\ILz{ }dynamics, thereby guaranteeing invariance of the
circulation of $\mathbf{\bar{u}}_{i}$ around $ \partial D$.  This
has been shown \citep[][]{Ripa-GAFD-93} to have important consequences
for the generalized Hamiltonian structure of the \ILz.

\subsection{A formulation suggestive of a generalized Hamiltonian
structure\label{Hamilton}}

The Euler equations of fluid mechanics possess what is called a
generalized Hamiltonian structure \citep[e.g.,][]{Morrison-82}. The
\ILi{ }(\ref{ILinf}), which derive from the Euler equations, are
also Hamiltonian in a generalized sense \citep[e.g.,][]{Abarbanel-etal-86}.
A good sign of the validity of any approximate model derived from
the \ILi{ }is the preservation of the generalized Hamiltonian
structure. This section is devoted to show that the $n$-\ILo{ }admits
a formulation suggestive of a generalized Hamiltonian structure. A
stronger statement was made in R95 for $1$-\ILo.

Let $\varphi(\x,t) = (\varphi ^{1}(\x,t),\dotsc ,\varphi^{7n}(\x,t))$
be a \textquotedblleft point\textquotedblright\ on the infinite-dimensional
phase space with coordinates $(\bar{\vartheta }_{i},\vartheta
_{i}^{\sigma },h_{i},\allowbreak \mathbf{\bar{u}}_{i}, \u_{i}^{\sigma
})$, $i=1,\dotsc,n$. Consider the relevant class, say $\mathfrak{A},$
of sufficiently smooth real-valued functionals of $ \varphi .$ For
any phase functional $\F[\varphi ]\in \mathfrak{A}$ it is further
assumed that its density does not depend explicitly on $t,$ namely,
$\F[\varphi ]=\int_{D}F(\varphi , \mathbf{\nabla }\varphi ,\allowbreak
\dotsc ,\x) \d{}^{2}\x $, and that it satisfies the boundary
conditions\footnote{ The symbol $\smash{\frac{\delta\F}{\delta
\varphi}}$ denotes the functional (variational) derivative of
$\F[\varphi]$, which is the unique element satisfying $\lim_{\varepsilon
\rightarrow 0}\varepsilon ^{-1}( \F[\varphi +\varepsilon \delta
\varphi ]\allowbreak -\F [\varphi ])=\int_{D}\smash{\frac{\delta\F}{\delta
\varphi}}\delta\varphi \d{}^{2}\x$ for arbitrary $\delta \varphi$.}
\begin{equation}
  \frac{\delta \F}{\delta \mathbf{\bar{u}}_{i}}\cdot \mathbf{\hat{n}}
  = 0 = \frac{\delta \F}{\delta \u_{i}^{\sigma }}\cdot \mathbf{
  \hat{n}}\quad \text{on}\quad \partial D.  
  \label{adm}
\end{equation}
A phase functional $\F[\varphi ]\in \mathfrak{A}$ will be said to
be \emph{admissible}. Introduce then the functional
\begin{equation}
  \H[\varphi ]:=\int_{j}E_{j},
\end{equation}
where
\begin{equation}
  \int_{j}\, := \int_{D}\d{}^{2}\x\sum_{j}
\end{equation}
and $E_i$ is the energy in the $i$th layer \eqref{eq:Ej}; its functional
derivatives are given by
\begin{equation}
  \frac{\delta \H}{\delta
  \bar{\vartheta}_{i}}=h_{i}(\tilde{h}_{i-1}+
  \tfrac{1}{2}h_{i}),\quad\frac{\delta \H}{\delta \vartheta
  _{i}^{\sigma }}=-\frac{h_{i}^{2}}{6},\quad\frac{\delta \H}{\delta
  h_{i}}=h_{i}\bar{b }_{i},\quad\frac{\delta \H}{\delta
  \mathbf{\bar{u}}_{i}}=h_{i}\mathbf{ \bar{u}}_{i},\quad\frac{\delta
  \H}{\delta \u_{i}^{\sigma }}= \frac{h_{i}\u_{i}^{\sigma}}{3}.
\end{equation}
The latter and the zero normal flow conditions across $\partial D$
(\ref{BC}) show that $\H$ is admissible. Let now
\begin{equation}
  \mathbb{J}=\bigoplus_{j} \mathbb{I}_{j}+\mathbb{K}_{j}
\end{equation}
be a skew-adjoint $7\times 7$ block-diagonal matrix operator where
$\mathbb{I}_{i}$ and $\mathbb{K}_{i}$ are expressed, for convinence,
in the following condensed form: 
\renewcommand{\arraystretch}{1.25}
\renewcommand{\arraycolsep}{.075in}
\begin{subequations}
\begin{equation}
  \mathbb{I}_{i}=-\left( 
  \begin{array}{ccccc}
  0 & 0 & 0 & 0 & 0 \\ 
  0 & 0 & 0 & 0 & 0 \\ 
  0 & 0 & 0 & \mathbf{\nabla }\cdot (\bullet ) & 0 \\ 
  0 & 0 & \mathbf{\nabla }(\circ ) & \bar{q}_{i}\mathbf{\hat{z}}\times
  (\bullet ) & q_{i}^{\sigma }\mathbf{\hat{z}}\times (\bullet ) \\ 
  0 & 0 & 0 & q_{i}^{\sigma }\mathbf{\hat{z}}\times (\bullet ) & 3\bar{q}_{i}
  \mathbf{\hat{z}}\times (\bullet )
  \end{array}
  \right),
\end{equation}  
\begin{equation}
  \mathbb{K}_{i}=-\left( 
  \begin{array}{ccccc}
  0 & 0 & 0 & h_{i}^{-1}(\bullet )\cdot \mathbf{\nabla }\bar{\vartheta}_{i} & 
  h_{i}^{-1}\mathbf{\nabla }\cdot \vartheta _{i}^{\sigma }(\bullet ) \\ 
  0 & 0 & 0 & h_{i}^{-1}(\bullet )\cdot \mathbf{\nabla }\vartheta _{i}^{\sigma
  } & 3h_{i}^{-1}(\bullet )\cdot \mathbf{\nabla }\bar{\vartheta}_{i} \\ 
  0 & 0 & 0 & 0 & 0 \\ 
  -h_{i}^{-1}(\circ )\mathbf{\nabla }\bar{\vartheta}_{i} & -h_{i}^{-1}(\circ )
  \mathbf{\nabla }\vartheta _{i}^{\sigma } & 0 & 0 & h_{i}^{-1}\u
  _{i}^{\sigma }\mathbf{\nabla }\cdot (\bullet ) \\ 
  \vartheta _{i}^{\sigma }\mathbf{\nabla (}h_{i}^{-1}\circ) & 
  -3h_{i}^{-1}(\circ )\mathbf{\nabla }\bar{\vartheta}_{i} & 0 & \mathbf{\nabla
  (}h_{i}^{-1}\u_{i}^{\sigma }\cdot \bullet) & 0
  \end{array}
  \right).
\end{equation}
\end{subequations}
Here, the circle (resp., bullet) in parenthesis indicates operation on a
scalar (resp., two-component vector). 
\renewcommand{\arraystretch}{1}
\renewcommand{\arraycolsep}{.03in}
Define further a bracket operation $\{\mathcal{\,},\}:\mathfrak{A}\times 
\mathfrak{A}\rightarrow \mathfrak{A}$ as 
\begin{equation}
  \{\F,\G\}:=\int_{D} \frac{\delta 
  \F}{\delta \varphi }\mathbb{J}\frac{\delta \G}{\delta
  \varphi}\d{}^{2}\x
  \label{PB}
\end{equation}
$\forall \F,\G[\varphi ]\in \mathfrak{A}$. Then
the layer model equations (\ref{IL1n}) can be written in the form
\begin{equation}
  \partial_{t}\varphi =\{\varphi ,\H\},  \label{H}
\end{equation}
which is equivalent to $\mathcal{\dot{F}}=\{\F,\H\}$
$ \forall \F[\varphi ]\in \mathfrak{A}.$ 

The bracket operator (\ref{PB}) satisfies $\{\F,\G\} = -\{\G,\F\}$
(anticonmmutativity), $\{\F, a\G + b\K\} = a\{\F, \G\} + b\{\F,\K\}$
(bilinearity), and $\{\mathcal{FG}, \K\} = \F\{\G,\mathcal{ K}\} +
\G\{\F,\K\}$ (Leibniz' rule),  where $a,b$ are arbitray numbers and
$\F,\G, \K[\varphi]$ are any admissible functionals. The anticonmutativity
property follows from the skew-adjointness of the matrix operator
$\mathbb{J} $ [boundary terms cancel out by virtue of (\ref{adm})].
The bilinearity property and Leibniz' rule are direct consequences
of the bracket's definition.

That system (\ref{IL1n}) can be cast in the form (\ref{H}) appear
to suggest that the $n$-\ILo{ }is Hamiltonian in a generalized
sense, with the functional $\H$ and the bracket operator
$\{\,,\}$ being the Hamiltonian and Poisson bracket, respectively.
However, the bracket (\ref{PB}) does not seem to qualify as Poisson
since $\{\{\F,\G\},\K\} + \{\{\G,
\K\},\F\} + \{\{\K,\F\},\G\}
= 0$ (Jacobi's identity) does not seem to hold.

In addition to independence of the choice of phase space coordinates,
the Hamiltonian structure conveys other important properties like
the direct linkage of conservation laws with symmetries via Noether's
theorem \citep[cf., e.g.,][]{Shepherd-90}. While the $n$-\ILo{
}cannot be proved to be Hamiltonian, its energy, $\H$, and $-\M$,
where $\M[\varphi]:=\int_{j}M_{j}$ is the zonal momentum of the
system, do appear to be generators of $ t$- and $x$-translations
because of (\ref{H}) and $\partial _{x}\varphi =\{\M,\varphi\}$,
repectively.  The latter assumes that $\M$ is an admissible functional,
which requires the horizontal domain to be $x$-symmetric since
$\smash{\frac{\delta \M}{\delta \u_{i}^{\sigma }}}\equiv \mathbf{0}$
and $\smash{\frac{\delta \M}{\delta \bar{v}_{i}}}\equiv 0,$ but
$\smash{\frac{\delta \M}{\delta \bar{u}_{i}}}=\gamma h_{i}\neq 0$.
Then $\delta _{\mathcal{H }}\H=\varepsilon \{\H,\H\}=\varepsilon
\dot{\H}\equiv 0$ for the infinitesimal variation $\delta _{\H
}\varphi :=\varepsilon \{\varphi ,\H\}=\varepsilon \partial _{t}\varphi
$ induced by $\H$ and $\delta _{\M}\H =\varepsilon \{\H,\M\}=-\varepsilon
\dot{\M} =-\varepsilon \int_{j}h_{j}\bar{\vartheta}_{j}\partial
_{x}h_{0}\equiv 0$ iff $\partial _{x}h_{0}\equiv 0$ for the
infinitesimal variation $\delta _{ \M}\varphi :=\varepsilon \{\varphi
,\M\}=-\varepsilon \partial _{x}\varphi $ induced by $\M$. Consequently,
conservation of $\H$ and $\M$ are linked, respectively, to $t$- and
$x$ -symmetries of $\H$ (horizontal domain and topography in this
case included).

A distinguished feature of generalized Hamiltonian systems is the
existence of Casimirs $\C[\varphi ]\in \mathfrak{A}$ which satisfy
$\{\C,\F\}\equiv 0$ $\forall \F[\varphi ]\in \mathfrak{A}$. The
Casimirs are thus integrals of motion, yet not related to (explicit)
symmetries because $\{\varphi , \C\}\equiv 0$ ($\C$ does not generate
any transformation).  The $i$th-layer integrals of volume, mass,
and buoyancy variance are all addmissible functionals that communte
with any admissible functional in the bracket in \eqref{PB}.  The
$n$-\ILo{ }does not seem to support additional ``Casimir'' invariants.

The possibility of deriving a stochastic $n$-\ILo{ }using the SALT
approach \citep{Holm-15} is constrained to the existence of a Kelvin
circulation theorem, which is lacking for the $n$-\ILo.  The lack
of a Kelvin circulation theorem is tied to the nonexistence of a
generalized Hamiltonian structure and associated Euler--Poincare
variational formulation for the $n$-\ILo.  While buiding parameterizations
of unresolved submesoscale motions does not seem plausible using
this flow-topolgy-preserving framework, investigating the contribution
of the submesoscale motions to transport at mesoscales is still
possible via direct numerical simulation.  For this the apparent
generalized Hamiltonian formualtion of the $n$-\ILo{ }can be helpul,
as finite-difference schemes that preserve the conservation laws
of the system might be sought using the bracket approach developed
in \citet{Salmon-04}.

\subsection{Arnold stability\label{Arnold}}

In R95 it was shown that a state of rest (or a steady state with
at most a uniform zonal current) in the $1$-\ILo{ }can be shown to
be formally stable using Arnold's \citeyearpar{Arnold-65,Arnold-66}
method if and only if (\ref{CondTheta}) is satisfied, i.e., if and
only if the buoyancy is everywhere positive and increases (resp.,
decreases) with depth within a layer with the rigid bottom (resp.,
rigid lid). Arnold's method for proving the stability of steady
solution of a system consists in searching for conditions that
guarantee the sign-definiteness of a general invariant which is
quadratic to the lowest-order in the deviation from that state; the
resulting conditions are only sufficient \citep[e.g.,][]{Holm-etal-83,
McIntyre-Shepherd-87}. In the $n$-\ILo{ }with $n>1$, however,
Arnold's method fails to provide stability conditions even for a
state of rest and with no topography ($h_{0}\equiv 0$ ). The
lowest-order (quadratic) contribution to that invariant, which can
be called a ``free energy'' because it is defined with respect to
a state of rest,
\begin{align}
  \mathcal{E} :={}& \frac{1}{2}\int_{j}H_{j}(\delta
  \mathbf{\bar{u}}_{j})^{2}+\tfrac{1}{3}H_{j}(\delta \u_{j}^{\sigma
  })^{2}+\left(g_{j}-\tfrac{1}{2}N_{j}^{2}H_{j}\right)(\delta h_{j})^{2}\nonumber\\ 
  &+N_{j}^{-2}H_{j}\left(\delta \bar{\vartheta}_{j}+\tfrac{1}{2}N_{j}^{2}\delta
  h_{j}\right)^{2}+\tfrac{1}{3}N_{j}^{-2}H_{j}(\delta \vartheta _{j}^{\sigma
  } - \tfrac{1}{2}N_{j}^{2}\delta h_{j})^{2}\nonumber\\ 
  &+ \left(g_{j}\delta h_{j}+H_{j}\delta
  \bar{\vartheta}_{j}\right)\delta \tilde{h}_{j-1},
\end{align}
cannot be proved sign-definite when $n>1$. Here, $H_{i},$ $g_{i}$
and $N_{i}$ are the $i$th-layer unperturbed depth, vertically
averaged buoyancy, and Brunt--V\"{a}is\"{a}l\"{a} frequency,
respectively. Similarly, a state of rest in the $n$-\ILz{ }for any
$n$ cannot be proved formally stable using Arnold's method.
Surprisingly, it is possible to prove the stability of a steady
state with a uniform zonal current in that model. But the condition
of stability is not one of ``static'' stability like (\ref{CondTheta})
as in the $1$-\ILo.  Contrarily, it is one of ``baroclinic'' stability
since a uniform current in the $n$-\ILz{ }has an implicit vertical
shear through the thermal-wind balance. These results can all be
inferred from \cite{Ripa-GAFD-93} and \cite{Ripa-JGR-96}.

Nevertheless, there is at least a system, which has one \ILz-like
layer and $n-1$ HL-like layers, for which a state of rest can be
proved formally stable. For instance, choosing the uppermost layer
to be \ILz-like, the corresponding free energy takes the form
\begin{align}
  \mathcal{E} :={}& \tfrac{1}{2}\int_{j}H_{j}(\delta \mathbf{
  \bar{u}}_{j})^{2}+\tfrac{1}{3}H_{\alpha }(\delta \u_{\alpha
  }^{\sigma })^{2}\nonumber\\ &+\tfrac{1}{2}N_{\alpha }^{-2}H_{\alpha
  }\left(\delta \bar{\vartheta}_{\alpha }+ \tfrac{1}{2}N_{\alpha
  }^{2}\delta h_{\alpha }\right)^{2 }+ \tfrac{1}{3}N_{\alpha
  }^{-2}H_{\alpha }\left(\delta \vartheta _{\alpha }^{\sigma
  }-\tfrac{1}{2} N_{\alpha }^{2}\delta h_{\alpha }\right)^{2}\nonumber\\
  &+(g_{j}-g_{j+1})(\delta \tilde{h}_{j})^{2}-\tfrac{1}{2}N_{\alpha
  }^{2}H_{\alpha }(\delta h_{\alpha })^{2},
\end{align}
where $\alpha :=n$ (resp., $\alpha :=1$) for the rigid-bottom (resp.,
rigid-lid) configuration, and $H_{i},$ $g_{i}$, and $N_{i}$ are all
constants. The above free energy is positive-definite if and only
if (\ref{CondTheta}) if fulfilled. [The $n$-HL has an infinite set
of invariants which are given by $\int_{j}h_{j}F(\bar{q}_{j})$ where
$ F(\cdot )$ is arbitrary; these include the volume integral, which
is the only one needed to obtain the above result.] When all layers
are homogeneous the same result is obtained. When one \ILz-like
layer is included, however, the free energy cannot be shown of one
sign.

That a steady state (with or without a current) of the $n$-\ILo{
}cannot be proved formally stable does not mean that such a state
is unstable; it actually means that Arnold's method is not useful to
provide sufficient conditions for the stability of that state.

\subsection{Waves\label{waves}}

The $n$-\ILo{ }equations (\ref{IL1n}), linearized with respect to
a reference state with no currents, can be shown to sustain the
usual midlatitude and equatorial gravity and vortical waves
(Poincar\'{e}, Kelvin, Rossby, Yanai, etc.) in $2n$ vertical normal
modes. Here I shall concentrate on how well these modes are represented
by considering the phase speed of (internal) long gravity waves
assuming a rigid-lid setting.

The reference state is characterized by the parameter
\begin{equation}
  S :=
  \frac{N_{\mathrm{r}}^{2}H_{\mathrm{r}}}{2g_{\mathrm{r}}},
\end{equation}
which must be such that $ 0<S<1 $ \citep[][]{Ripa-JFM-95,Beron-Ripa-97}.
Here, $N_{\mathrm{r}}$ is the reference Brunt--V\"{a}is\"{a}l\"{a}
frequency within an active layer floating on top of an inert layer; $H_{%
\mathrm{r}}$ is the total thickness of the active fluid layer; and $g_{%
\mathrm{r}}$ denotes the vertically averaged reference buoyancy
within the active layer. All three reference quantities are held
constant. The reference buoyancy then varies linearly from
$g_{\mathrm{r}}(1+S)$ at the top of the active layer to
$g_{\mathrm{r}}(1-S)$ at the base of the active layer. In R95 it
was shown that the $1$-\ILo{ }gives the exact result for the
``equivalent'' barotropic or external mode phase speed of (internal)
long gravity waves for all $S$, and a very good approximation to
the first internal mode phase speed for all $S$.

\begin{figure}[t!]
  \centering
  \includegraphics[width=.75\textwidth]{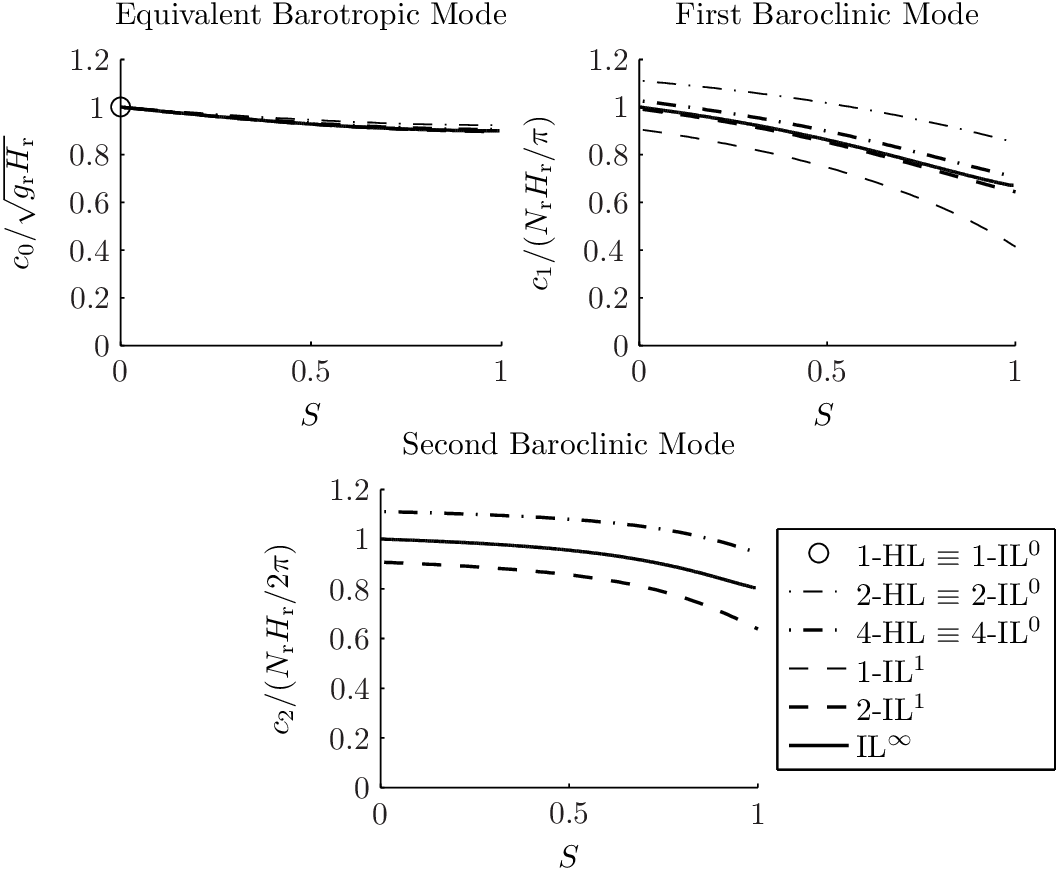}%
  \caption{Phase speed of (internal) long gravity waves as a function
  of the stratification strength in a reduced-gravity reference state
  with no currents.}
  \label{vermod}%
\end{figure}

\figref{vermod} compares, as a function of $S$, the phase speed as
determined by the \ILi, $n$-HL, $n$-\ILz, and $n$-\ILo{ }for various
$n$.  The figure shows the results for the external mode ($c_{0}$),
and the first ($c_{1}$) and second ($c_{2}$) internal modes. The
analytical expression for the \ILi's phase speed for an arbitrary
mode number can be found in R95; the phase speeds for the layer
models are computed numerically. The solutions of the $n$-HL and
$n$-\ILz{ }coincide because $\bar{\vartheta}_i$ is constant for a
normal mode in the $n$-\ILz.  These models can only support $n$
vertical normal modes. In contrast, the $n$-\ILo{ }sustains vertical
normal modes up to the $(n+1)$th internal mode.

As noted above, the 1-\ILo{ } result coincides with that of \ILi{
}for the barotropic mode.  To approximate well the exact solution,
two HL- or \ILz-like layers are needed.  The first internal mode
solution is very well approximated using two \ILo-like layers.  Four
HL-like layers do not provide a similar degree of approximation.
The second internal mode solution is reasonably approximated with
two \ILo-like layers. The distance between the exact solution and
that produced using four HL-like layers is of the same order.
However, in every case the $n$-HL (or the $n$-\ILz) overestimates
the exact phase speeds.

\subsection{Baroclinic instability}\label{instability}

As one further test of the validity of the $n$-\ILo, the problem
of baroclinic instability, particularly upper-ocean baroclinic
instability, is considered here. (A subset of the results presented
here appeared in \citet{Beron-etal-04b}.) The behavior in both
quasigeostrophic and ageostrophic regimes is explored. The $n$-\ILo{
}solutions are compared in all cases with the \ILi{ }solutions. In
some cases comparisons are also made with $n$-HL and $n$-\ILz{
}solutions.  In the quasigeostrophic regime analytical expressions
exist for the \ILi{ }solutions. Analytical or semianalytical formulas
for the dispersion relations also exist in this regime for the
1-\ILo{ }and models with one \ILz-like or two HL-like layers. The rest
of the solutions shown are computed numerically upon finite
differencing the corresponding eigenvalue problems.

Upper-ocean baroclinic instability, e.g., above the ocean thermocline,
is studied in \citet[]{Beron-Ripa-97} using the \ILi{ }and the
1-\ILo{ } in a reduced-gravity setting.  A basic state with a
parallel current $\U = U(z)\,\hat\x$ is considered in that work to
lie in an infinite channel on the $f$ plane, to have a uniform
vertical shear, and to be in thermal-wind balance with the
across-channel buoyancy gradient.  The basic velocity is further
set to vary (linearly) from $\bar{U}+U^{\sigma }$ at the top of the
active layer to $\bar{U}-U^{\sigma }$ at the base of the active
layer. Accordingly, the basic buoyancy field $\Theta(y,z)$ varies
from $g_{\mathrm{r}}(1-2fU^{\sigma }y/H_{\mathrm{ r}}+S)$ at the
top of the active layer to $g_{\mathrm{r}}(1-2fU^{\sigma
}y/H_{\mathrm{r}}-S)$ at the base of the active layer ($y$ is the
across-channel coordinate).  A nonvanishing velocity at the base
of the active layer implies that the latter has a linear $y$-slope
given by $g_{\mathrm{r}}^{-1}f\left( U^{\sigma }-\bar{U}\right)/(1-S)$.
This basic state is a steady solution of the \ILi{ }to the lowest
order in the Rossby number, $\mathrm{Ro} := \bar U/L|f| \sim
U^\sigma/L|f|$ where $L$ is the relevant length scale, which is
assumed to be an infinitesimal parameter. In the limit of weak
stratification $(S\rightarrow 0)$ the horizontal scales
\begin{equation}
  R_{\mathrm{E}}:=\frac{\sqrt{g_{\mathrm{r}}H_{\mathrm{r}}}}{\left|
  f\right|},\quad
  R_{\mathrm{I}}:=\frac{N_{\mathrm{r}}H_{\mathrm{r}}}{\left|
  f\right|}
\end{equation}
are well separated $(R_{\mathrm{E}}\gg R_{\mathrm{I}})$, and thus
long and short normal-mode perturbations to this state can be
identified. Under long small-Rossby-number normal-mode perturbations
the base of the active layer behaves as a free boundary. For short
small-Rossby-number normal-mode perturbations this interface is
effectively rigid. When the vertical shear is assumed strong,
$\bar{U}/U^{\sigma }\ll O(S^{-1}),$ the short-perturbation limit
corresponds to the classical Eady problem of baroclinic instability,
in whose case solutions are insensitive to $\bar{U} /U^{\sigma}$.

\begin{figure}[t!]
  \centering 
  \includegraphics[width=.45\textwidth]{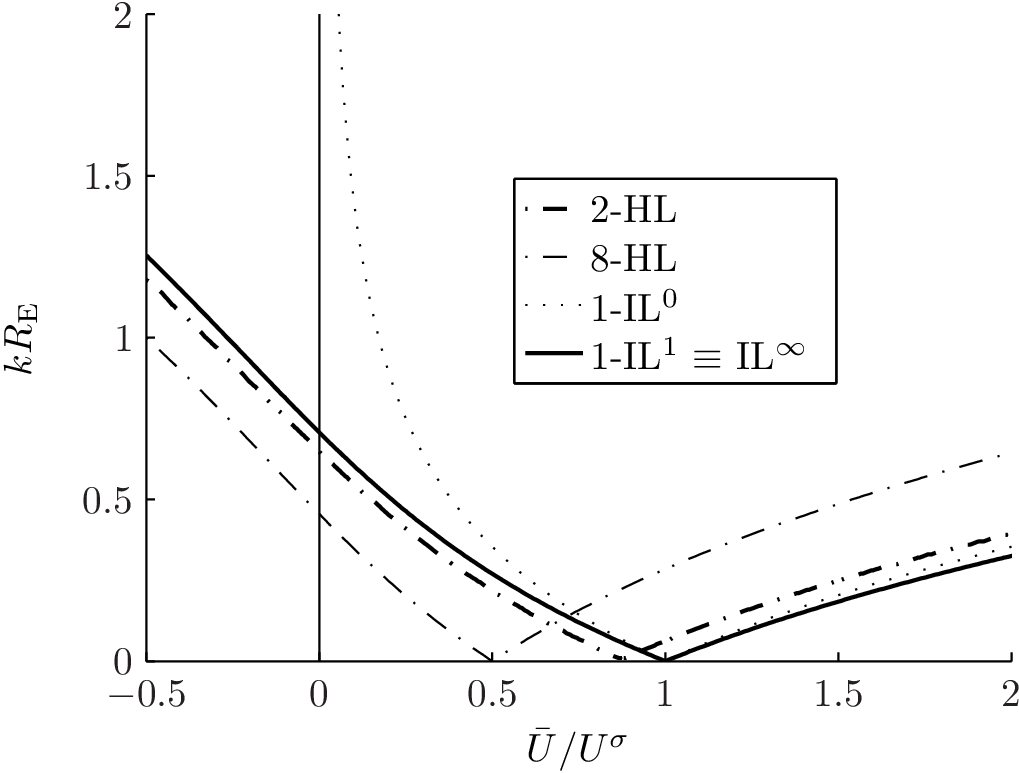}\quad
  \includegraphics[width=.45\textwidth]{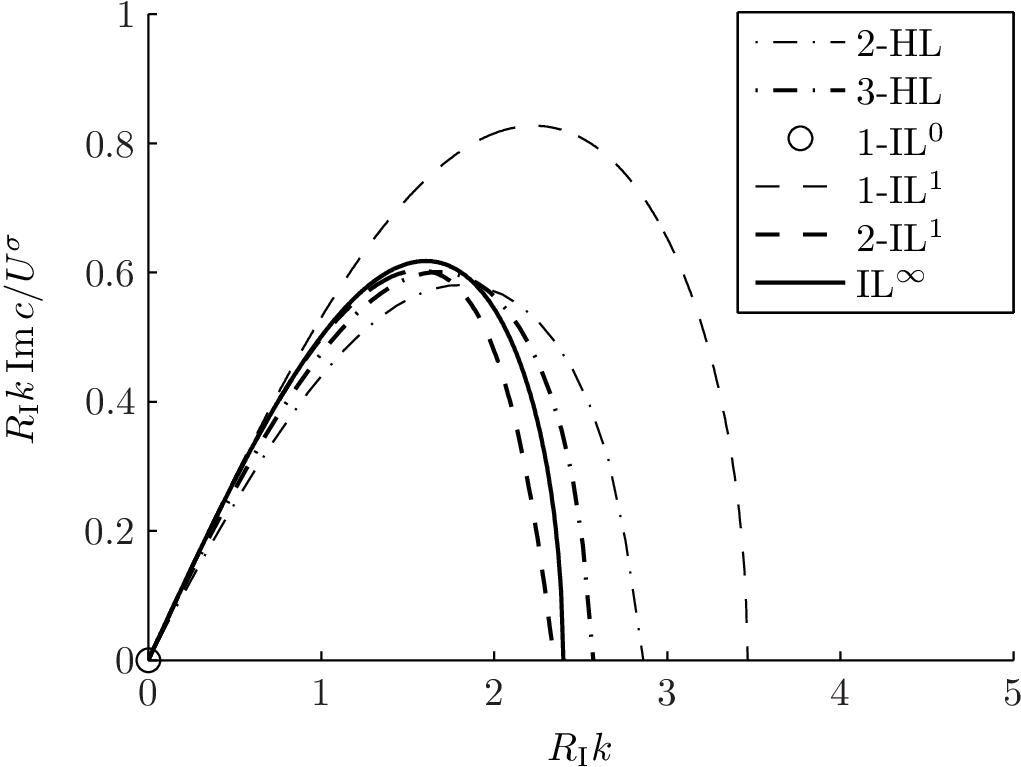}%
  \caption{(left panel) Minimum wavenumber for long-perturbation and
  strong-shear (i.e., free-boundary) baroclinic instability as a
  function of the slope of the lower interface in the basic state.
  (right panel) Growth rate of the most unstable perturbation as a
  function of the wavenumber in short-perturbation, strong-shear
  (i.e., classical Eady) baroclinic instability.} 
  \label{barinst}%
\end{figure}

The left panel of \figref{barinst} shows the minimum along-channel
wavenumber, $k$, for instability as a function of $\bar{U}/U^{\sigma
}$ in the long-perturbation and strong-shear limits (free-boundary
baroclinic instability). The 1-\ILo{ }gives the exact result for
all $\bar{U}/U^{\sigma}$ \citep[][]{Beron-Ripa-97}. To provide a
close approximation to this result for all $\bar{U}/U^{\sigma}$
with the $n$-HL, a fairly large $n$ (cir.\ $25$) is needed.  Note
that the 1-\ILz{ }predicts, incorrectly, stability for $\bar{U}/U^{\sigma
}<0$ (the vertical shear in this model is implicit through the
thermal-wind relation).

The right panel of \figref{barinst} depicts, as a function of the
along-channel wavenumber $k$, the growth rate of the most unstable
perturbation in the short-perturbation and strong-shear limits
(classical baroclinic instability). The comparison of the
maximum growth rate predicted by the 1-\ILo{ }with the \ILi's maximum
growth rate is less satisfactory in this limit. However, and very
importantly, a high wavenumber cutoff of baroclinic instability is
present.  The 1-\ILz{ } model only gives the $k=0$ value of the
growth rates of
this figure and thus it cannot be used to describe this regime ($N_{\mathrm{r%
}}\equiv 0$ in this model). Three \ILo-like layers are enough to
approximate well the exact maximum growth rate for all $k$. To
obtain a similar result using HL-like layers, at least $6$ must be
considered.

\begin{figure}[t!]
  \centering
  \includegraphics[width=.45\textwidth]{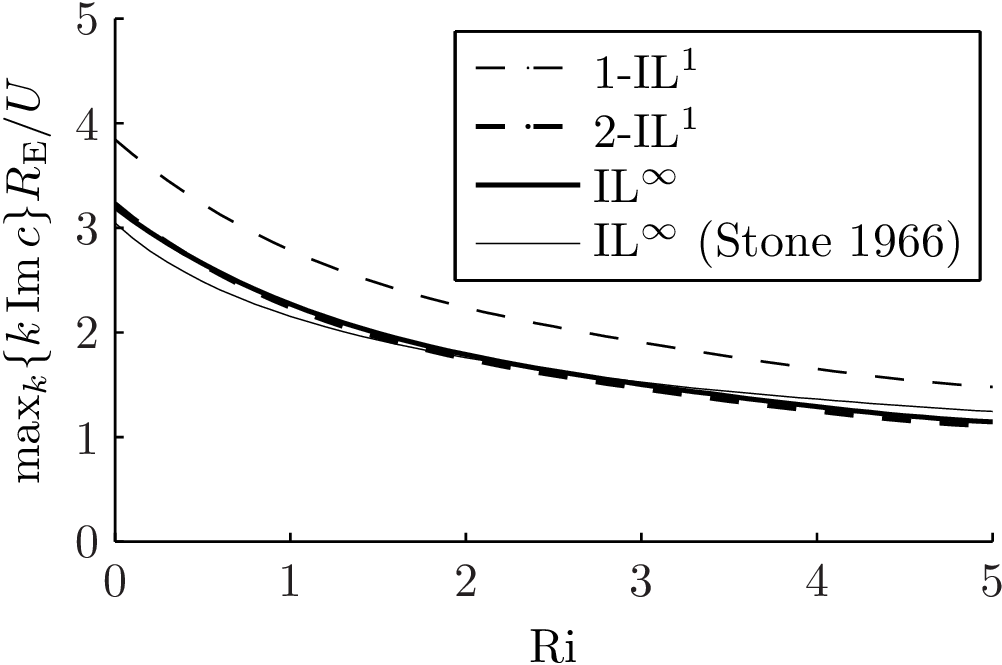}\quad
  \includegraphics[width=.45\textwidth]{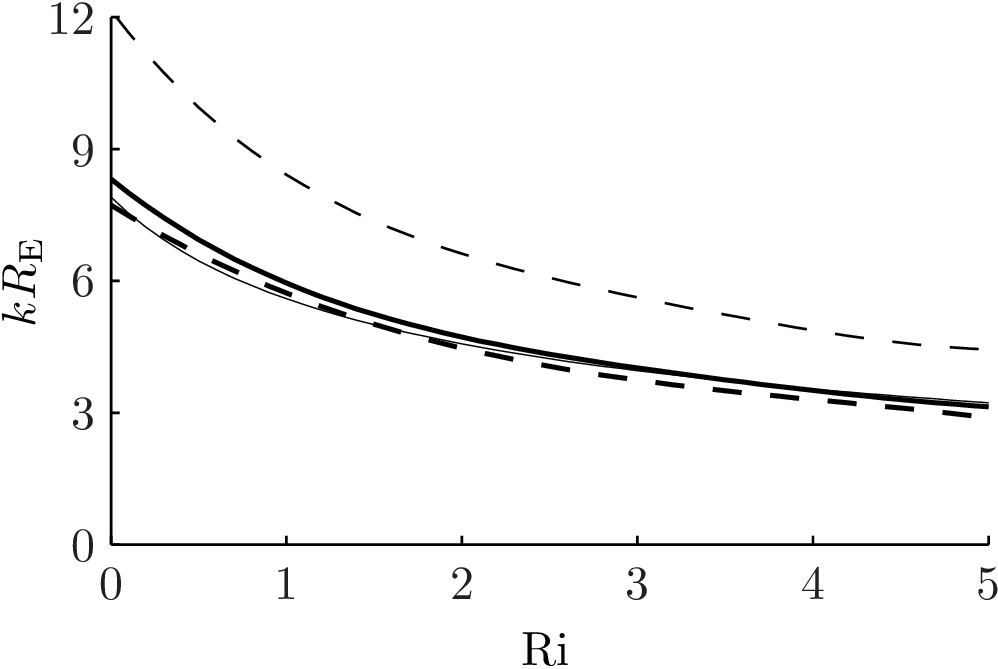}%
  \caption{(left panel) Maximum normal-mode perturbation growth rate
  for ageostrophic (classical Stone) baroclinic instability as a
  function of the Richardson number. (right panel) Wavenumber for
  maximum growth rate.}
  \label{IL1n-Stone}%
\end{figure}

For the basic state considered above the Richardson number
\begin{equation}
  \mathrm{Ri} := \left(\frac{N_\mathrm{r}}{\partial_z U}\right)^2
  \equiv \frac{SR_\mathrm{E}^{2}}{2\mathrm{Ro}^{2}L^{2}}.
\end{equation}
In classical baroclinic instability, for which $\mathrm{Ro},S\rightarrow
0$, $L=R_{ \mathrm{I}}\equiv \sqrt{2S}R_{\mathrm{E}}$ and
$\bar{U}/U^{\sigma }\ll O(S^{-1}),$ the well-known result
$\mathrm{Ri}\rightarrow \infty $ holds. In free-boundary baroclinic
instability, for which $\mathrm{Ro}, S\rightarrow 0$, $L=R_{\mathrm{E}}$
and $\bar{U}/U^{\sigma }\ll O(S^{-1}),$ $ \mathrm{Ri}$\textrm{\
}can acquire any value because a proper way \citep[][]{Beron-Ripa-97}
to achieve the $S\rightarrow 0$ limit is to set $S=O(\mathrm{Ro}^{\nu
})$ for any $\nu$. Unlike quasigeostrophic baroclinic instability,
ageostrophic baroclinic instability is characterized by a dependence
of the solutions on $ \mathrm{Ri}$ \citep[][]{Stone-66,Stone-70}.
This dependence is checked in the present layer model by considering
infinitesimal nongeostrophic normal-mode perturbations to the above
basic state but with $\bar{U}\equiv U^{\sigma }=U$, and assuming
$\mathrm{Ro} =10^{-1}$ and $L=R_{\mathrm{E}}$.

The left panel of \figref{IL1n-Stone} shows, as a function of
$\mathrm{Ri}$, the maximum growth rate $\max_{k}\{k\Im c\}$ of the
perturbation. The right panel of the figure shows, also as a function
of $\mathrm{Ri}$, the wavenumber, $k_{\max}$, at which the latter
value is attained. Shown for reference is an \ILi{ }asymptotic
solution, valid up to $O(\mathrm{Ro }^{3})$. The asymptotic formulas
for $\max_{k}\{\Im kc\}$ and $k_{\max }$ are those given in equations
(4.27) and (4.28) of \citet[]{Stone-66}.  The $n$-\ILo{ }fares very
well even with $n=1$.  A model with a single \ILz-like layer,
however, cannot describe this regime because of the dependence on
$\mathrm{Ri}$ (for the 1-\ILz{ }$S\equiv 0$).  With two \ILo-like
layers the maximum growth rates and corresponding wavenumbers at
which they are achieved are in very close agreement with the \ILi{
}predictions in the range of $\mathrm{Ri}$ values explored, which
was much wider than that shown in \figref{IL1n-Stone}.  Note,
however, that observations indicate that typical values of $\mathrm{Ri}$
in the upper ocean are close to unity \citep[e.g.,][]{Tandon-Garrett-94}

\subsection{Forcing\label{forcing}}

In R95 forcing (wind stress, interfacial drag, and buoyancy/heat
input) was introduced in the $1$-\ILo{ }model equations in a way
that was compatible with the conservation laws of energy, momentum,
and mass/heat content. The same approach is adopted here to include,
in addition, freshwater fluxes through the surface in accordance
with the conservation law of salt content. The possibility for the
exchange of fluid across the other interfaces is also considered.

Let $\mathbf{\tau }(\x,t)$ be a wind stress acting at the surface
of the ocean ($\rho _{n+1}\equiv 0$ must be the setting in the
rigid-bottom configuration and typically $h_{0}\equiv 0$ in the
rigid-lid one). Assume further that there is a friction force acting
at the interface between contiguous layers. Introduction of these
forces in Newton's equations (\ref{IL1n}d,e) in the form
\begin{subequations}\label{tau}
\begin{align}
  \partial _{t}\mathbf{\bar{u}}_{i}+\cdots &=\delta _{i\alpha }\mathbf{\tau }
  /h_{\alpha }-r_{i}(\mathbf{\bar{u}}_{i}\pm \u_{i}^{\sigma }),\\
  \partial _{t}\u_{i}^{\sigma }+\cdots &=\mp 3\delta _{i\alpha }
  \mathbf{\tau }/h_{\alpha }+3r_{i}(\mathbf{\bar{u}}_{i}\pm \u
  _{i}^{\sigma }),
\end{align}
\end{subequations}
implies that the work done by the wind stress is proportional to
the velocity at the top of the uppermost layer, $\mathbf{\bar{u}}_{\alpha
}\mp \u_{\alpha }^{\sigma}$, and that one done by the friction force
in the $i$th layer is proportional to the velocity at the base of
that layer, $\mathbf{\bar{u}}_{i}\pm \u_{i}^{\sigma}$.  Namely,
\begin{subequations}
\begin{align}
  \partial _{t}\sum_{j}E_{j}+\cdots &=\mathbf{\tau }\cdot (\mathbf{\bar{u}}
  _{\alpha }\mp \u_{\alpha }^{\sigma })-\sum_{j}r_{j}h_{j}(\mathbf{
  \bar{u}}_{j}\pm \u_{j}^{\sigma })^{2},\\
  \partial _{t}\sum_{j}M_{j}+\cdots &=\mathbf{\tau }\cdot \mathbf{\hat{x}}
  -\sum_{j}r_{j}h_{j}(\mathbf{\bar{u}}_{j}\pm \u_{j}^{\sigma })\cdot 
  \mathbf{\hat{x}}.
\end{align}
\end{subequations}
In the above equations, $\delta_{ij}$ is the Kroenecker delta and
$r_{i}$ is a friction coefficient that can be taken as a constant
or as some function of $h_{i}$ and $|\mathbf{\bar{u}}_{i}\pm
\u_{i}^{\sigma }|.$ [Recall that $\alpha :=n$ (resp., $\alpha :=1$)
for the rigid-bottom (resp., rigid-lid) configuration.]

Let now $\Gamma (\x,t)$ be a buoyancy input through the surface and
write the buoyancy equations (\ref{IL1n}a,b) in the
form
\begin{subequations}\label{b&h}
\begin{align}
  \partial _{t}\bar{\vartheta}_{i}+\cdots &=\delta _{i\alpha }\Gamma
  /h_{\alpha }, \\
  \partial _{t}\vartheta _{i}^{\sigma }+\cdots &=\eta \delta _{i\alpha
  }\Gamma /h_{\alpha },
\end{align}
where $\eta $ is any constant. Consider, in addition, the possibility
of fluid crossing the interface between consecutive layers; then
the volume conservation equation (\ref{IL1n}c) can be rewritten as
\begin{equation}
  \partial _{t}h_{i}+\cdots =w_{i}^{\mathrm{b}}-w_{i}^{\mathrm{t}}.
  \end{equation}
\end{subequations}
Here, the quantities $w_{i}^{\mathrm{t}}(\x,t)$ and $w_{i}^{\mathrm{b
}}(\x,t)$ are volume fluxes per unit area through the top and base
of the $i$th layer, respectively. The set (\ref{b&h}), for any value
of $ \eta$, is compatible with the mass conservation equation
\begin{equation}
  \partial _{t}(h_{i}\bar{\vartheta}_{i})+\cdots =\delta _{i\alpha }\Gamma +
  \bar{\vartheta}_{i}(w_{i}^{\mathrm{b}}-w_{i}^{\mathrm{t}}).
\label{masaforzada}
\end{equation}
At the surface $w_{\alpha }^{\mathrm{t}}(\x,t)=E(\x,t)-P( \x,t),$
which represents the imbalance of evaporation minus
precipitation.\footnote{More precisely, $w_{\alpha
}^{\mathrm{t}}=(1-s)(E-P)\approx E-P$ with $s( \x,t)$ being the
salt fraction (salinity times $10^{-3}$) at the surface
\citep{Beron-etal-99}.} Away from the surface some parametrization
must be adopted. In models with \ILz-like layers it is commonly set
\citep[e.g.,][]{McCreary-etal-91}
\begin{equation}
  w_{i}^{\mathrm{b}}-w_{i}^{\mathrm{t}}=(-1)^{i+1}\dfrac{(h_{i-1}-H_{i-1}^{
  \mathrm{e}})^{2}}{H_{i-1}^{\mathrm{e}}t_{i}^{\mathrm{e}}}\theta
  (H_{i-1}^{ \mathrm{e}}-h_{i-i}).
\end{equation}
Here, $H_{i}^{\mathrm{e}}$ and $t_{i}^{\mathrm{e}}$ are constants
that with units of length and time, respectively, that characterize
the ``entrainment'' process, and $\theta(\cdot )$ is the Heaviside
step function. In the present case, an algorithm may be designed
such that condition (\ref{CondTheta}) is fulfilled at all times.
This would allow for a more natural representation of mixing
processes, including the possibility of representing localized
mixing events, e.g., characterized by $\bar{\vartheta}_{i+1}+\vartheta
_{i+1}^{\sigma }<\bar{\vartheta} _{i}-\vartheta _{i}^{\sigma }$
instantaneously at certain position.  This subject deserves to be
studied in detail.

Let finally assume a linear state equation, i.e., $\vartheta
_{i}=g\alpha _{T}(T_{i}-T_{n+1})-g\alpha _{S}(S_{i}-S_{n+1}).$ Here,
$\alpha _{T}$ and $ \alpha _{S}$ are the thermal expansion and salt
contraction coefficients, respectively; $T_{i}(\x,\sigma
,t)=\bar{T}_{i}(\x,t)+\sigma T_{i}^{\sigma }(\x,t)$ and $S_{i}(\x,\sigma
,t)=\bar{S}_{i}( \x,t)+\sigma S_{i}^{\sigma }(\x,t)$ are the $i$th
layer temperature and salinity, respectively; and $T_{n+1}$ and
$S_{n+1}$ are the inactive layer (constant) temperature and salinity,
respectively. Let also write the buoyancy input as
\begin{equation}
  \Gamma =g\alpha _{T}(\rho _{\mathrm{r}}C_{p})^{-1}Q+g\alpha
  _{S}\bar{S} _{\alpha }(P-E),  \label{split}
\end{equation}
$C_{p}$ is the specific heat at constant pressure and $Q(\x,t)$ is
the heat input through the surface. Equation (\ref{masaforzada}) can then be
split into a heat and salt content conservation equations,
namely, 
\begin{subequations}
\begin{align}
  \partial _{t}(h_{i}\bar{T}_{i})+\cdots &=\delta _{i\alpha }(\rho
  _{\mathrm{r
  }}C_{p})^{-1}Q+\bar{T}_{i}(w_{i}^{\mathrm{b}}-w_{i}^{\mathrm{t}}), \\
  \partial _{t}(h_{i}\bar{S}_{i})+\cdots &=\delta _{i\alpha
  }\bar{S}_{\alpha
  }(E-P)+\bar{S}_{i}(w_{i}^{\mathrm{b}}-w_{i}^{\mathrm{t}}).
\end{align}
\end{subequations}
If fluid across the surface is allowed only, the choice (\ref{split})
enforces, on one hand \citep[e.g.,][]{Beron-etal-99},
\begin{subequations}\label{HS}
\begin{equation}
  \frac{\mathrm{d}}{\mathrm{d}t}\int_{j}h_{j}\bar{S}_{j}\equiv 0,
\end{equation}
and, on the other \citep[][]{Beron-Ripa-00},
\begin{equation}
  \frac{\mathrm{d}}{\mathrm{d}t}\left\langle T\right\rangle =V^{-1}\int_{D}
  (\rho _{\mathrm{r}}C_{p})^{-1}Q \d{}^2\x + (\bar{T}_{\alpha
  }-\left\langle T\right\rangle)(P-E),
\end{equation}
\end{subequations}
where $V:=\int_{j}h_{j}\equiv \int_{D}\mathrm{d}^{2}\x$ $h$ is the
total volume and $\left\langle T\right\rangle :=V^{-1}\int_{j}h_{j}\bar{T}
_{j}$ is the average temperature in $V$. Note that (\ref{HS}\emph{b}),
unlike the equation satisfied by $\int_{j}h_{j}\bar{T}_{j},$ is
independent---as it should---of the choice of the origin of the
temperature scale \citep[cf.][]{Warren-99}.

\section{Concluding remarks\label{discussion}}

This paper describes a multilayer extension of the single-layer
primitive-equation model for ocean dynamics and thermodynamics
introduced in \citet{Ripa-JFM-95}. Inside each layer the velocity
and buoyancy fields can vary not only arbitrarily in the horizontal
position and time, but also linearly with depth.

In the absence of external forcing and dissipation, the model
conserves volume, mass, buoyancy variance, energy, and zonal momentum
for zonally symmetric horizontal domains and topographies. Unlike
models with depth-independent velocity and buoyancy fields within
each layer, the model generalized here is able to represent the
thermal wind balance explicitly at low frequency inside each layer.
In this sense, the model of this paper has ``better'' physics than
a model with depth-independent fields.  For a fixed number of layers,
the model of this paper can sustain one more vertical normal mode
than the homogeneous-layer models, which, on the other hand, are
not able to incorporate thermodynamic processes (e.g., due to heat
and buoyancy fluxes across the air--sea interface or associated
with localized vertical mixing events). In this other sense, the
present model has ``more'' physics than a model with homogeneous
layers. Last but not least, overall improved results in both
quasigeostrophic (free-boundary and classical Eady) and ageostrophic
(classical Stone) baroclinic instability with respect to the
single-layer calculations are attained with the addition of a small
number layers.

The present generalization enriches Ripa's single-layer model by
providing it enough flexibility to approach problems for which a
single-layer structure is too idealized. Configurations with a small
number of layers are particularly useful for the insight they provide
into physical processes.  Configurations with more layers may provide
the basis for an accurate numerical circulation model.

Finally, and returning to the motivation for revisiting the
construction of models with reduced thermodynamics, the requirement
on the two-dimensional structure of the models is satisfied by the
model derived here.  A different strategy than that taken here is
needed to fulfill the requirement on the geometric structure of the
models, if the goal is to pursue flow-topology-preserving
parameterizations of unresolved scales using the SALT (stochastic
advection by Lie transport) framework  \citep{Holm-15, Holm-Luesink-20}.
The desired result might follow from plugging Ripa's ansatz in the
Hamilton principle's Lagrangian of the primitive equations for
continuously stratified fluid.  This is currently under investigation.
A stochastic parameterization framework that can be applied to the
model derived here is location uncertainty (LU) \citep{Resseguier-etal-20}.
Unlike SALT dynamics, which preserve Kelvin circulation, the LU
framework conserves energy, so it can be immediately applied on the
present model and is a natural fit to considering the parameterizations
based on extraction of available potential energy
\citep{Gent-McWilliams-90, Fox-etal-08, Bachman-etal-17}.  Building
stochastic parameterizations using the generalized Ripa's model is
left for future work.

\paragraph{Acknowledgements.} 

A stimulating epistolary exchange with Darryl Holm provided incentive
to revisit this work and finish it.

\bibliographystyle{mybst-notitle} 

\end{document}